\documentclass[aps,prd,showpacs,twocolumn,superscriptaddress,nofootinbib,preprintnumbers]{revtex4-1} 

\makeatletter
\def\p@subsection{}
\makeatother

\usepackage{color,graphicx,amsmath,amssymb,lipsum}
\usepackage[colorlinks=true,citecolor=blue,linkcolor=blue,urlcolor=blue, backref=false,pdfborder={0 0 0}]{hyperref}

\usepackage[normalem]{ulem}
\usepackage[utf8]{inputenc} 
\usepackage{mathtools}
\usepackage{wrapfig}

\usepackage[table,svgnames,dvipsnames]{xcolor}
\usepackage{booktabs}

\definecolor{vlightgray}{gray}{0.9}
\usepackage{float}

\usepackage{ulem}
\usepackage{diagbox}

\usepackage{bm}
\usepackage{bbm}

\renewcommand\thesection{\arabic{section}}

\newcommand{\be}{\begin{equation}}
\newcommand{\ee}{\end{equation}}
\newcommand{\beqa}{\begin{eqnarray}}
\newcommand{\eeqa}{\end{eqnarray}}

\renewcommand\k{ {\bf k}}

\newcommand{\fnl}{f_{\rm NL}}

\newcommand{\hMpc}{h \text{Mpc}^{-1}}
\newcommand{\q}{{\bf q}}

\newcommand{\bseq}{\begin{subequations}}
\newcommand{\eseq}{\end{subequations}}

\renewcommand{\ln}{\mathop{\rm ln}\nolimits}

\def\gsim{\raise0.3ex\hbox{$\;>$\kern-0.75em\raise-1.1ex\hbox{$\sim\;$}}}
\def\lsim{\raise0.3ex\hbox{$\;<$\kern-0.75em\raise-1.1ex\hbox{$\sim\;$}}}

\def\beqn#1{\begin{equation}\label{#1}}
\def\eeqn{\end{equation}}

\def\beqa#1{\begin{eqnarray}\label{#1}}
\def\eeqa{\end{eqnarray}}

\def\Z2{$\mathcal{Z_2}$}

\newcommand {\ignore}[1]{}

\begin{document}


\preprint{CERN-TH-2022-005}

\title{Constraints on Single-Field Inflation 
from the BOSS Galaxy Survey}

\author{Giovanni Cabass}
\email{gcabass@ias.edu}
\affiliation{School of Natural Sciences, Institute for Advanced Study, 1 Einstein Drive, Princeton, NJ 08540, USA}

\author{Mikhail M. Ivanov}
\thanks{Einstein Fellow}
\email{ivanov@ias.edu}
\affiliation{School of Natural Sciences, Institute for Advanced Study, 1 Einstein Drive, Princeton, NJ 08540, USA}

\author{Oliver H.\,E. Philcox}
\affiliation{Department of Astrophysical Sciences, Princeton University, Princeton, NJ 08540, USA}
\affiliation{School of Natural Sciences, Institute for Advanced Study, 1 Einstein Drive, Princeton, NJ 08540, USA}

\author{Marko Simonovi\'c}
\affiliation{Theoretical Physics Department, CERN, 1 Esplanade des Particules, Geneva 23, CH-1211, Switzerland}
\author{Matias Zaldarriaga}
\affiliation{School of Natural Sciences, Institute for Advanced Study, 1 Einstein Drive, Princeton, NJ 08540, USA}

\begin{abstract} 
\noindent Non-local primordial non-Gaussianity (NLPNG) is
a smoking gun of interactions 
in single-field inflationary models, 
and can be written as a combination of the equilateral and orthogonal 
templates. 
We present the first constraints 
on these
from the redshift-space 
galaxy power
spectra and bispectra of the Baryon
Oscillation Spectroscopic Survey (BOSS) data. 
These are the first 
such measurements independent of the cosmic microwave background fluctuations.
We perform a consistent 
analysis that
includes all necessary nonlinear corrections 
generated by NLPNG,
and vary all relevant cosmological and nuisance parameters 
in a global fit to the data.
Our conservative analysis
yields joint limits on the amplitudes 
of the equilateral and orthogonal shapes,
$\fnl^{\rm equil}=940\pm 600$, $\fnl^{\rm ortho}=
-170\pm 170$ (both at 68\%~CL).
These can be used to derive 
constraints on coefficients of the effective
single-field inflationary Lagrangian; in particular, we find that the 
sound speed of inflaton fluctuations has the bound $c_s\geq 0.013$
at 95\%~CL. 
Fixing the 
quadratic galaxy bias and cosmological
parameters, 
the constraints 
can be tightened to $\fnl^{\rm equil}=260\pm 300$, $\fnl^{\rm ortho}=
-23\pm 120$ (68\%~CL).
\end{abstract}

\maketitle

\textit{Introduction ---} Cosmology is the interface between particle physics 
and general relativity. Nothing exemplifies this more than inflation -- 
a primordial accelerated
expansion of the Universe
that may have happened 
at energy scales as high as $10^{14}\,{\rm GeV}$.
Inflation naturally 
generates
quantum 
fluctuations 
that provide the seeds for
the formation and clustering of matter
and galaxies.
Thus, observations of the large-scale structure of our Universe allow us to probe
physics at these extremely high energies, inaccessible to present-day particle accelerators.

There are three main questions about inflation one may ask:
What is its energy scale?
How many degrees of freedom generated density fluctuations?
How fast did these degrees of freedom propagate?
While significant efforts have been devoted to answering the first question, by constraining the amplitude of primordial 
gravitational waves, the latter two require a probe of deviations of the initial density 
fluctuations from 
a Gaussian distribution,
known as primordial non-Gaussianity (PNG).

The simplest observable encoding PNG is the bispectrum, $B_{\zeta}$,
of the primordial metric curvature perturbation $\zeta$. 
Due to translational and rotational invariance,  $B_{\zeta}$ 
is a function of the moduli of three momenta, $\k_1,\k_2,\k_3$, which form a closed triangle. 
A bispectrum peaking at squeezed triangles, $k_1\ll k_2\approx k_3$, is a generic signature of particle 
interactions in multi-field inflation~\cite{Linde:1996gt,Enqvist:2001zp,Lyth:2001nq,Lyth:2002my,Dvali:2003em,Dvali:2003ar,Zaldarriaga:2003my,Byrnes:2010em,Senatore:2010wk},\footnote{These shapes also appear in
ekpyrotic alternatives to inflation (see Ref.~\cite{Lehners:2010fy} for a review), 
but they typically produce strong PNG 
incompatible with data~\cite{Planck:2019kim}.} i.e.~where more than one degree of freedom 
is light during inflation. 
This type of PNG is 
called ``local''.
In contrast, a bispectrum peaking 
at equilateral ($k_1\approx k_2\approx k_3$)
or flattened ($k_1\approx k_2\approx k_3/2$) triangles 
is a peculiar feature of 
interactions in single-field inflation~\cite{Arkani-Hamed:2003juy,Alishahiha:2004eh,Senatore:2004rj,Chen:2006nt,Creminelli:2006xe,Cheung:2007st,Cheung:2007sv,Senatore:2009gt},
which has only one 
degree of freedom (inflaton). 
This kind of 
``non-local'' primordial non-Gaussianity (NLPNG)
can be represented as a linear 
combination of two basis shapes,
equilateral and orthogonal~\cite{Senatore:2009gt}, with amplitudes $\fnl^{\rm equil}$ and $\fnl^{\rm ortho}$ respectively.

Symmetries of inflation also 
dictate a relationship 
between the 
inflaton 
speed of sound 
and the strength 
of nonlinear interactions that generate NLPNG~\cite{Cheung:2007st}.
In particular, there is a theorem stating 
that
PNG can be large if and only if the sound speed is small~\cite{Baumann:2015nta,Green:2020ebl}. 
This allows one to constrain the propagation speed
of the inflaton from the observed level of NLPNG.

Up to now, the only source of information on NLPNG
has been the cosmic microwave 
background 
(CMB) temperature and polarization 
data~\mbox{\cite{2013ApJS..208...20B,Planck:2019kim}}. 
In particular, \emph{Planck} 2018 data yield $\fnl^{\rm equil}=-26\pm 47$, $\fnl^{\rm ortho}=
-38\pm 24$ (both at 68\%~CL)~\cite{Planck:2019kim}. In theory,
one can obtain better constraints
with upcoming galaxy surveys,
which will collect orders of magnitude more cosmological information 
as counted
in number of accessible Fourier modes.
However, the analysis
of this data is also more
intricate because  PNG in
the galaxy distribution is 
a weak effect on top of an intrinsic, late-time
non-Gaussian signal generated by nonlinear 
clustering of matter. Thus, late-time nonlinearities act 
as background noise, which must be
accurately modeled if we are to 
measure PNG from large-scale structure data.

There have been significant efforts to probe 
\textit{local} PNG from galaxy surveys, exploiting the scale-dependent bias that enhances 
the observed power spectrum on very large scales~\cite{Leistedt:2014zqa,Castorina:2019wmr,Mueller:2021tqa}. 
This enhancement originates from a particular 
form of the squeezed limit of the local shape.
NLPNG is different as 
the relevant shapes are suppressed in the  
squeezed limit and hence
do not produce  significant scale-dependent bias. 
Thus, this type of PNG requires a 
dedicated study.
Indeed, the leading effect of NLPNG on galaxy clustering
is a specific shape dependence in the 
galaxy bispectrum, which also modulates power spectra through loop
corrections. These effects can be 
constrained by a systematic analysis based on the 
consistently-modeled power spectrum 
and bispectrum data.
In this \textit{Letter}, we present the first such analysis of the publicly
available state of the art redshift clustering data 
from the  Baryon  Oscillation  Spectroscopic  Survey  (BOSS).

A rigorous analysis of the galaxy 
bispectrum has been 
a long-time challenge. 
Even ignoring PNG, the complete theoretical model including 
all necessary effects relevant for the actual data
 has been developed only recently~\cite{Ivanov:2021kcd,Philcox:2021kcw}~(see also~\cite{Scoccimarro:2000sn,Sefusatti:2006pa,Baldauf:2014qfa,Angulo:2014tfa,Eggemeier:2018qae,Ivanov:2019pdj,DAmico:2019fhj,Oddo:2019run,Eggemeier:2021cam,Alkhanishvili:2021pvy,Oddo:2021iwq,Chen:2021wdi,Baldauf:2021zlt}).
On the data side, a major improvement has come 
from the estimators that remove 
effects of the survey window function~\cite{Philcox:2020vbm,Philcox:2021ukg}.
These efforts now allow us to
obtain the first CMB-independent 
limits on NLPNG from 
galaxy redshift surveys.

\textit{PNG in single-field inflation 
---} 
A general single-field Lagrangian
for the inflaton perturbation $\pi$
with leading interactions up to cubic order
is given by~\cite{Cheung:2007st,Senatore:2009gt}
\be
\label{eq:eft}
\begin{split}
& S_{\rm EFT}=\int {\rm d}^4x \sqrt{-g}\Bigg[
{-\frac{M_{\rm P}^2\dot{H}}{c_s^2}}\left(\dot\pi^2 -c_s^2\frac{({\bm\nabla}\pi)^2}{a^2}\right)\\
&+\frac{M_{\rm P}^2\dot{H}}{c_s^2}(1-c_s^2)
\left(\frac{\dot{\pi}({\bm\nabla}\pi)^2}{a^2}-
\left(1+\frac{2}{3}\frac{\tilde{c}_3}{c_s^2}\right)\dot{\pi}^3
\right)
\Bigg]\,\,.
\end{split}
\ee
$\pi$ is
related to the metric curvature perturbation via $\zeta=-H\pi$, with $H$ being the Hubble 
parameter during inflation.
The two inflaton interactions are parametrized by
the sound speed $c_s$ and a dimensionless Wilson 
coefficient $\tilde{c}_3$. It is customary to represent 
PNG produced by these interactions as a linear combination of 
the orthogonal
and equilateral templates.
To that end, we define 
\begin{equation} 
\label{eq:B_primordial_master_definition}
B_\zeta(k_1,k_2,k_3) = \frac{18}{5}f_{\rm NL}\Delta^4_\zeta\frac{{\cal S}(k_1,k_2,k_3)}{k_1^2k_2^2k_3^2}\,\,. 
\end{equation}
Here, $\Delta^2_\zeta$ is the amplitude 
of the primordial power spectrum: 
$k^3P_\zeta(k)=\Delta^2_\zeta (k/k_\ast)^{n_s-1}$, where $n_s$ is the spectral index.
Analysis of \emph{Planck} data finds $\Delta^2_\zeta\approx 4.1\times 10^{-8}$, $n_s\approx 0.96$~\cite{Aghanim:2018eyx} for the pivot scale $k_\ast=0.05$ Mpc$^{-1}$.
The equilateral and orthogonal templates are defined as~\cite{Babich:2004gb,Senatore:2009gt}\footnote{Note that we use the orthogonal template from Appendix B of \cite{Senatore:2009gt}, 
which has a physically 
correct suppression 
in the squeezed limit $k_1\ll k_2\approx k_3$, where
it goes as $\mathcal{S}_{\rm ortho}\propto k_1/k_3$. 
It can be
contrasted with the commonly used approximate template that 
features an unphysical enhancement in that limit~\cite{Planck:2019kim}.
} 
\begin{equation}
    \begin{split}
       & {\cal S}_{\rm equil}(k_1,k_2,k_3) =  \\
&
\bigg(\frac{k_1}{k_2} + \text{$5$ perms.}\bigg)- \bigg(\frac{k_1^2}{k_2k_3} + \text{$2$ perms.}\bigg) - 2\,\,, \label{eq:equilateral_template}  \\
& {\cal S}_{\rm ortho}(k_1,k_2,k_3) =(1+p)\,\frac{\Delta}{e_3} - p\,\frac{\Gamma^3}{e_3^2}\,\,,
    \end{split}
\end{equation}
where $p = 8.52587$, $\Delta = (k_T-2k_1)(k_T-2k_2)(k_T-2k_3)$, 
\be 
\begin{split}
    & k_T = k_1+k_2+k_3\,\,,\quad e_2=k_1k_2+k_2 k_3 + k_1 k_3\,\,,\\
    &e_3 = k_1 k_2 k_3\,\,,\quad 
    \Gamma = \frac{2}{3}e_2 - \frac{1}{3}(k_1^2+k_2^2+k_3^2)\,\,.
\end{split}
\ee 

The amplitudes 
$f_{\rm NL}^{\rm equil}$, $f_{\rm NL}^{\rm ortho}$ are related to  the cofficients
of the 
EFT Lagrangian via~\cite{Babich:2004gb,Senatore:2009gt} 
\be 
\label{eq:decomposition} 
\begin{split}
\begin{pmatrix} f_{\rm NL}^{\rm equil} \\
f_{\rm NL}^{\rm ortho}
\end{pmatrix}
 & = 
\begin{pmatrix}
1.04021 & 1.21041 \\
-0.0395140 & -0.175685
\end{pmatrix}
\begin{pmatrix}
f_{\rm NL}^{\dot{\pi}({\bm\nabla}\pi)^2} \\
f_{\rm NL}^{\dot{\pi}^3}
\end{pmatrix}\,\,, 
\\
f_{\rm NL}^{\dot{\pi}({\bm\nabla}\pi)^2} & =\frac{85}{324}(1-c_s^{-2})\,,\\
f_{\rm NL}^{\dot{\pi}^3}
&=\frac{10}{243}(1-c_s^{-2})
 \left(\tilde{c}_3+\frac{3}{2}c_s^2\right)\,.
\end{split}
\ee

\textit{Large Scale Structure in the presence of NLPNG ---} 
Before considering NLPNG, we first discuss structure formation 
in a Universe with Gaussian initial conditions.
We describe it in the framework of the effective field 
theory of large-scale structure (EFTofLSS),
where one builds a perturbative expansion 
in terms of the linear matter overdensity field $\delta^{(1)}$
~(see~\cite{Baumann:2010tm,Carrasco:2012cv,Desjacques:2016bnm,Ivanov:2019pdj,DAmico:2019fhj,Nishimichi:2020tvu}
and references therein).
For Gaussian initial conditions, statistical properties 
of $\delta^{(1)}$ are fully determined 
by its power spectrum\footnote{We suppress the explicit time dependence for
brevity.} $P_{11}$
\be
\langle  \delta^{(1)}(\k) \delta^{(1)}(\k') \rangle  = (2\pi)^3\delta^{(3)}_D(\k'+\k)P_{11}(k)\,\,.
\ee
We restrict our galaxy
power spectrum
analysis 
to 
the one-loop order in the EFTofLSS. In the absence of NLPNG, the nonlinear power spectrum takes the following schematic form
\be
P_{\rm G} \equiv  P_{11}+ P_{\rm 1-loop}+P_{\rm ctr}+P_{\rm stoch}\,\,,
\ee
where 
$P_{\rm 1-loop}$ is the 
nonlinear one-loop correction for Gaussian initial conditions, 
$P_{\rm ctr}$ is the counterterm that stems 
from the higher-derivative operators in the fluid equation
and the bias expansion, while $P_{\rm stoch}$ captures 
galaxy stochasticity.
Nonlinear clustering also generates 
a non-trivial ``Gaussian'' bispectrum $B_{\rm G}$,
which we consider here in the tree-level approximation.

PNG affects large-scale structure via three channels:
initial conditions, loop corrections, and 
scale-dependent galaxy bias.
First, PNG induces a bispectrum signal
additional to the one coming from nonlinear mode coupling~\cite{Sefusatti:2007ih,Sefusatti:2009qh},
\be 
\label{eq:inib}
\begin{split}
& \langle \delta^{(1)}\delta^{(1)}\delta^{(1)}\rangle = \fnl B_{111}(k_1,k_2,k_3) (2\pi)^3\delta^{(3)}_D(\k_{123})\,\,,\\
& \fnl B_{111}(k_1,k_2,k_3)=\mathcal{T}(k_1)\mathcal{T}(k_2)\mathcal{T}(k_3)B_\zeta(k_1,k_2,k_3)\,\,,
\end{split}
\ee
where $\k_{123}=\k_1+\k_2+\k_3$, and we introduced the transfer functions 
$\mathcal{T}(k)\equiv \delta^{(1)}(\k)/\zeta(\k)=(P_{11}(k)/P_\zeta(k))^{1/2}$. 
The initial bispectrum \eqref{eq:inib} also
leaks into the nonlinear galaxy power 
spectra through 
mode coupling, and
induces the following non-Gaussian one-loop correction (which adds to $P_{\rm G}$ above)
\be 
\label{eq:P12_expression}
\begin{split}
& \fnl P_{12}(\textbf{k})  = 2\fnl Z_1(\k)
\\
& \times
\int\frac{{\rm d}^3q}{(2\pi)^3} Z_2(\q,\k-\q) B_{111}(k,q,|\k-\q|)\,\,.
\end{split}
\ee
Here $Z_{1,2}$ are linear and quadratic galaxy redshift-space kernels~\cite{Chudaykin:2020aoj},
$Z_1(\k)=b_1+f 
(\hat{\k}\cdot \hat{\mathbf{z}})^2$
where $f$ is the logarithmic growth factor, $\hat{\mathbf{z}}$ is the line-of-sight direction unit vector, and $\hat{\k}\equiv \k/k$.
The $P_{12}$ terms are present for both galaxies and matter~\cite{Taruya:2008pg,Assassi:2015jqa,MoradinezhadDizgah:2020whw}, 
and we will refer to them as ``NG matter loops''
in what follows.

Third, NLPNG 
modulates galaxy formation,
which is captured on large scales 
by
the  
scale-dependent galaxy bias~\cite{Assassi:2015fma,Angulo:2015eqa,Desjacques:2016bnm},\footnote{Equilateral and orthogonal PNG produce identical 
scale-dependent biases that can be captured by a single free parameter.}
 \be
 \label{eq:bphi}
 \delta_g = b_1\delta +f_{\rm NL}b_\zeta (k/k_{\rm NL})^2 \zeta + \text{nonlinear}\,\,,
 \ee 
 where $\delta$ and $\delta_g$ 
 are overdensity fields of matter
 and galaxies respectively, 
 $b_1$ is the usual linear bias, $b_\zeta$
 is an order-one PNG linear bias
 coefficient, 
 and
 $k_{\rm NL}\approx 0.5\,\hMpc$ is the nonlinear scale\footnote{Defined by $P_{11}(k_{\rm NL})k^3_{\rm NL}/(2\pi^2)=1$.} 
 at the 
 relevant 
 redshift $z\simeq 0.5$.

The relative size of various perturbative corrections
can be estimated using the scaling universe approach~\cite{Pajer:2013jj,Assassi:2015jqa}.
It is based on the observation that the 
linear power spectrum 
can be well 
approximated by a power-law $P_{11}\propto (k/k_{\rm NL})^{n}k_{\rm NL}^{-3}$ with $n\approx -1.7$ for 
quasi-linear wavenumbers
$k\simeq 0.15\,\hMpc$. 
Assuming that there is a single nonlinear scale in the problem,
the scaling universe estimates suggest that 
the leading non-Gaussian corrections
are the PNG matter loops and the linear
scale-dependent bias. 
The total dimensionless galaxy power spectrum $\Delta^2(k)\equiv k^3 P(k)/(2\pi^2)$ can be estimated as
\be\begin{split}
\Delta^2(k)& =
{\underbrace{\left(\frac{k}{k_{\rm NL}}\right)^{1.3}}_{\rm tree}}+
\underbrace{\left(\frac{k}{k_{\rm NL}}\right)^{2.6}}_{\rm 1-loop} 
+ \underbrace{\left(\frac{k}{k_{\rm NL}}\right)^{3.3}}_{\rm ctr} 
+ \underbrace{\left(\frac{k}{k_{\rm NL}}\right)^{3}}_{\rm stoch}  \\
& +\underbrace{\fnl \Delta_\zeta\left(\frac{k}{k_{\rm NL}}\right)^{1.95}}_{\text{NG matter loops}}  
+\underbrace{\fnl b_{\zeta}\Delta_\zeta\left(\frac{k}{k_{\rm NL}}\right)^{2.65}}_{\text{linear PNG bias}}\,\,.
\end{split} 
\ee
All higher order
corrections, such as PNG terms $\mathcal{O}(\fnl^2)$,
NG two-loop corrections,
contributions generated by
nonlinear bias operators like $ 
\delta \nabla^2\zeta$, and Gaussian two-loop
corrections, 
are subleading
for $\fnl \Delta_\zeta \lesssim 0.1$ 
and $k\lesssim 0.17~\hMpc$
typical for our analysis, and hence can be neglected.\footnote{For local-type PNG 
the situation is different: the PNG bias 
terms dominate
on large scales.}
This will be validated
on the mock simulation data below. 
As for the bispectrum,
we use the tree-level approximation 
so only the leading PNG bispectrum \eqref{eq:inib} is of importance~\cite{Assassi:2015jqa}.

All in all, our final model for the galaxy power spectra
and bispectra in redshift space 
is given by
\be
\begin{split}
& P(\k) = P_{\rm G}(\k)+f_{\rm NL}\left(P_{12}(\k) 
+\frac{2 b_\zeta Z_1(\k) k^2}{k_{\rm NL}^2}
\frac{P_{11}(k)}{\mathcal{T}(k)}\right)\,,\\
& B(\k_1,\k_2,\k_3) =  B_{\rm G}(\k_1,\k_2,\k_3) \\
&~~~~+f_{\rm NL}Z_1(\k_1)Z_1(\k_2)Z_1(\k_3)B_{111}(k_1,k_2,k_3) 
\,\,,
\end{split}
\ee
where $ P_{\rm G}$
and $B_{\rm G}$ are the standard 
Gaussian power spectrum and bispectrum models~\cite{Ivanov:2019hqk,Ivanov:2021kcd}.
In practice, 
we compute the Legendre redshift-space multipoles 
$P_{\ell}$ ($\ell=0,2,4$) of the galaxy power spectrum
and use the angle-averaged (monopole) bispectrum.
We also implement IR resummation in redshift space~\cite{Senatore:2014via,Baldauf:2015xfa,Blas:2015qsi,Blas:2016sfa,Ivanov:2018gjr,Vasudevan:2019ewf} (to account for long-wavelength displacement effects) and the Alcock-Paczynski 
effect~\cite{Alcock:1979mp} (to account for coordinate conversions~\cite{Ivanov:2019pdj}).

Our model has 14 nuisance parameters: 13 standard
bias parameters and counterterms 
of Gaussian redshift-space power spectra 
and bispectra (present in previous analyses), plus the
scale-dependent PNG bias $b_\zeta$~\eqref{eq:bphi}.
,

\textit{Data and Analysis ---} 
We use the twelfth  data release  (DR12)~\cite{Alam:2016hwk} of  BOSS.   
The data is split into two redshift  bins with effective means $z=0.38,0.61$,  
in  each  of  the  Northern  and Southern galactic caps, resulting in 
four independent data chunks.
The survey contains $\sim 1.2\times 10^6$ galaxy  positions  with  a  total  volume  of  $6\,(h^{-1}\text{Gpc})^3$.
From each chunk, we  use  the  power  spectrum multipoles ($\ell=0,2,4$)
for $k\in [0.01,0.17)\,\hMpc$, the real-space power spectrum $Q_0$
for $k\in [0.17,0.4)\,\hMpc$ \cite{Ivanov:2021haa}, the redshift-space bispectrum monopoles
for triangle configurations within the range of $k_i\in [0.01,0.08) \,\hMpc$ (62 triangles),
and the BAO parameters
extracted 
from the post-reconstructed power spectrum data~\cite{Philcox:2020vvt}, as in~\cite{Philcox:2021kcw}. 
The power spectra and bispectra are measured with the window-free estimators~\cite{Philcox:2020vbm,Philcox:2021ukg}.
The covariances 
for each data chunk
are computed from a suite of 2048  MultiDark-Patchy  mocks~\cite{Kitaura:2015uqa}.

\begin{figure}[htb!]
\centering
\includegraphics[width=0.49\textwidth]{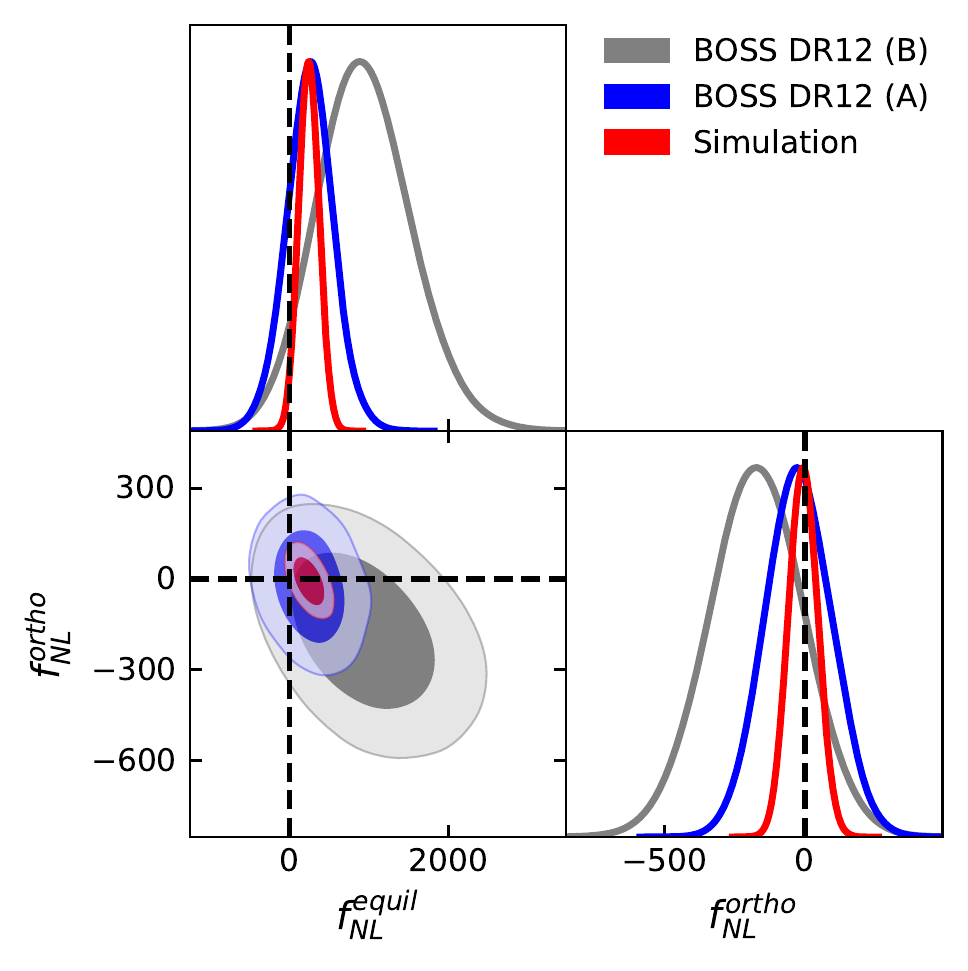}
\caption{
Marginalized constraints on parameters ($\fnl^{\rm equil},\fnl^{\rm ortho}$)
from the BOSS data obtained in 
the conservative baseline analysis
(BOSS DR12 (B), gray), and in the
aggressive analysis
(BOSS DR12 (A), blue).
We also show results
from the full Nseries simulation suite (red), whose
volume is 40 times larger 
than BOSS.
Dashed lines indicate $\fnl^{\rm equil}=0,\fnl^{\rm ortho}=0$.
\label{fig:fnl} }
\end{figure}

\begin{figure}[htb!]
\centering
\includegraphics[width=0.49\textwidth]{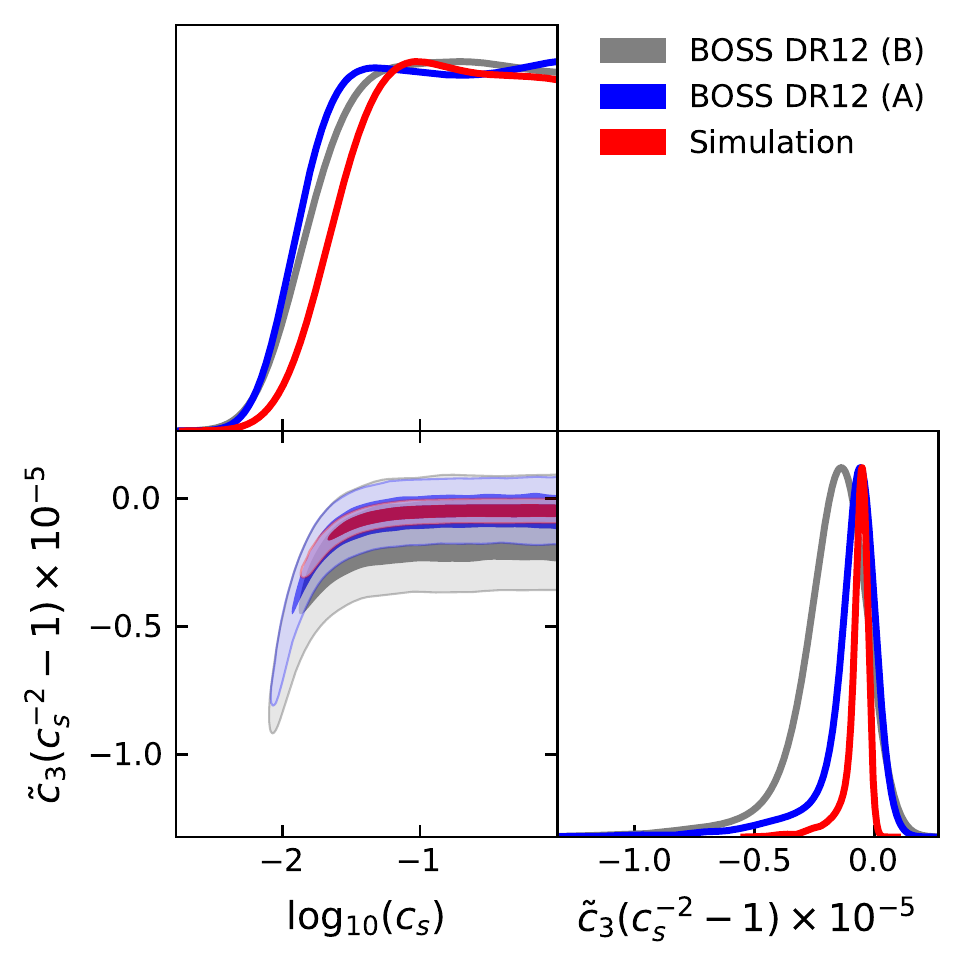}
\caption{Marginalized constraints on single-field
inflation parameters ($c_s,\tilde{c}_3$)
from the BOSS data in the 
baseline analysis (BOSS DR12 (B), gray), 
aggressive analysis (BOSS DR12 (A), blue), 
and from 
the full Nseries simulation suite (red). 
\label{fig:cs} } 
\end{figure}

We perform the full-shape analysis of the redshift
clustering data following the methodology of~\cite{Ivanov:2019pdj,DAmico:2019fhj,Philcox:2020vvt,Philcox:2021kcw}.
We implement the complete theory model for the
power spectra and bispectra of galaxies in redshift space 
in an extension of the \texttt{CLASS-PT} code~\cite{Chudaykin:2020aoj}
\footnote{Code available at 
\href{https://github.com/Michalychforever/CLASS-PT}{{github.com/Michalychforever/CLASS-PT}}, with custom MontePython likelihoods available at \href{https://github.com/oliverphilcox/full_shape_likelihoods}{github.com/oliverphilcox/full\_shape\_likelihoods}.}
that includes all non-Gaussian corrections described above (computed using the FFTlog approach~\cite{Simonovic:2017mhp}).
We consistently recompute the shape of these corrections as we scan over different
cosmologies in a Markov Chain Monte Carlo (MCMC) analysis. 
Up to additional NG contributions, our analysis is identical to~\cite{Philcox:2021kcw}.

In our baseline analysis we fix the baryon density to the BBN measurement~\cite{Cooke:2017cwo}, 
the primordial power spectrum tilt to the Planck best-fit value~\cite{Aghanim:2018eyx},
and the neutrino mass 
to the minimal value allowed by oscillation experiments $\sum m_\nu =0.06$~eV.
We vary 
the physical dark matter density $\omega_{cdm}$, 
the reduced Hubble 
parameter $h$, the amplitude of the primordial
scalar fluctuations $\ln (10^{10}A_s)$,
and 
$(\fnl^{\rm equil},\fnl^{\rm ortho})$
within flat infinitely wide priors.
We use the same priors for nuisance parameters as~\cite{Philcox:2021kcw}.
We also marginalize over
the linear PNG bias, 
$b_{\zeta} =   1.686 \cdot  \frac{18}{5} (b_1-1) \tilde{b}_{\zeta}$
within a Gaussian prior $\tilde{b}_{\zeta}\sim \mathcal{N}(1,5)$
motivated by the peak-background split model~\cite{Schmidt:2010gw}.

In addition, we perform a more aggressive analysis,
whereupon cosmological parameters are set to the \textit{Planck} 2018 priors.
Moreover, instead of marginalization, we fix the quadratic galaxy biases 
to predictions
of the standard 
dark matter 
halo relations~\cite{Desjacques:2016bnm}. These
agree with
simulations at the level
required by the BOSS data~\cite{Eggemeier:2021cam,Ivanov:2021kcd}.

It is worth noting that our analysis is 
different from the CMB one~\cite{Senatore:2009gt},
where $\fnl$ is estimated directly from the temperature 
and polarization maps. 
In contrast, we do a global fit 
to the summary statistics,  
and vary $\fnl^{\rm equil},\fnl^{\rm ortho}$
along with all other  
cosmological and nuisance parameters in our MCMC chains, 
necessary due to the appearance of parameter-dependent late-time non-Gaussianity.

\textit{Results ---} 
First, we apply our pipeline to 
Nseries mock catalogs. These 
are based on high-resolution N-body simulations and were used 
by the BOSS collaboration for validation
tests~\cite{Alam:2016hwk}. The cumulative volume of this 
simulation is $235\,(h^{-1}\text{Gpc})^3$, 
which is $\approx 40$ times larger than 
actual BOSS survey volume.
The mocks were generated
from purely Gaussian initial conditions, 
which we must recover with our pipeline as a consistency check. 
Indeed, we
find $\fnl^{\rm equil}=240_{-130}^{+130}$, 
$\fnl^{\rm ortho}=-6_{-54}^{+52}$, 
which is consistent with $\fnl^{\rm equil}=0$ within 95\%~CL
and $\fnl^{\rm ortho}=0$ within 68\%~CL.
The shift in $\fnl^{\rm equil}$,
even if statistically insignificant, 
yields an estimate of the theoretical uncertainty, 
$\Delta\fnl^{\rm equil}\lesssim 100$,
which is  less than $0.2\sigma$
of the BOSS 1d marginalized statistical error.
This is consistent
with estimates of 
higher order perturbative
corrections that 
are omitted in our
analysis.
We have also found 
that the constraints 
are driven by 
the bispectrum:
the power spectrum 
data alone gives
$\fnl^{\rm equil}=-640\pm 1200$, 
$\fnl^{\rm ortho}=2400\pm 1600$
(at 68\%~CL).

Having validated our method on the simulations,
we move to the actual BOSS data.
Our baseline MCMC analysis yields the following  
1d marginalized estimates of equilateral  and  orthogonal $\fnl$
values from the joint fit:
\be
\fnl^{\rm equil}=940_{-650}^{+570} \,\,,\quad \fnl^{\rm ortho}=
-170_{-170}^{+180}\,\,,\quad \text{(68\%~CL)}\,\,.
\ee
We do not find any evidence for NLPNG: the 95\%~CL limits read
\be
 -280<\fnl^{\rm equil} < 2190\,\,,\quad -520<\fnl^{\rm ortho} < 176\,\,.
\ee
The resulting 
posterior contours are shown in Fig.~\ref{fig:fnl}.
The correlation coefficient between the equilateral and orthogonal shapes is
$-0.40$. The best measured 
principal component is
\[
\fnl^{\rm ortho}-0.11\fnl^{\rm equil}=-65 \pm 157\,\,.
\]
Note that the correlation is dictated by 
degeneracy directions in the 
shape of 
the BOSS galaxy bispectrum.
This can be contrasted with 
the \textit{Planck} 2018 data
that does not show any appreciable 
degeneracy between the two shapes, 
i.e.\ their correlation matches the intrinsic 
template 
cosine~\cite{Planck:2019kim}. 
This implies that combinations
with the CMB data will be 
important for degeneracy 
breaking 
in future analyses.

Our aggressive analysis yields 
noticeably 
stronger constraints, 
$\fnl^{\rm equil}=257_{-300}^{+300}$, 
$\fnl^{\rm ortho}=-23_{-120}^{+120}$.
Most of the improvement here 
is driven by fixing quadratic 
bias parameters. 
In light of the well-known
$\sigma_8$ tension, we have also
varied $\sigma_8$ in our aggressive fit, 
which somewhat loosens the bounds,
$\fnl^{\rm equil}=510_{-440}^{+320}$, 
$\fnl^{\rm ortho}=-123_{-140}^{+150}$ 
($\sigma_8=0.728_{-0.037}^{+0.033}$), at 68\%~CL.
Concluding, 
we emphasize that our 
aggressive analysis 
is performed
mainly for the purposes 
of illustration. 
It 
demonstrates that
PNG constraints can be 
significantly improved with
a better knowledge 
of bias parameters,
which motivates 
further work on their calibration with 
simulations, e.g. 
along the lines of~\cite{Barreira:2021ukk,Lazeyras:2021dar}.

We have also repeated 
our baseline analysis with free $n_s$ and found 
consistent, albeit somewhat weaker limits 
$\fnl^{\rm equil}=1200_{-850}^{+630}$, 
$\fnl^{\rm ortho}=-240_{-180}^{+210}$ (with $n_s=0.83\pm 0.08$).

\textit{Constraints on early Universe physics ---} 
Our main analysis was performed using flat priors on $\fnl^{\rm equil},\fnl^{\rm ortho}$,
which map onto
non-flat priors of 
the 
physical coefficients in the EFT Lagrangian~\eqref{eq:eft}.
To avoid any potential prior bias, 
we repeat our MCMC analysis 
varying directly the relevant parameters 
$\log_{10} c_s$ and $\tilde c_3 (c_s^{-2}-1)$ (following~\cite{Senatore:2009gt}),
assuming flat priors $\log_{10} c_s\in (-\infty,0]$
and $\tilde c_3 (c_s^{-2}-1) \in (-\infty,+\infty)$.
The corresponding 
posterior contours are shown in Fig.~\ref{fig:cs}.
They yield the bound on the sound speed
\be
c_s \geq 0.013 \quad (\tilde{c}_3~\text{marginalized, 95\%~CL})\,\,.
\ee
For DBI
inflation \cite{Alishahiha:2004eh}, where $\tilde{c}_3=3(1-c_s^2)/2$,
we find $c_s\geq 0.04$ (95\%~CL).
In general, the parameter $\tilde{c}_3$
cannot be constrained 
because of degeneracy with $c_s$,
i.e.~$\tilde{c}_3$ is unbounded
in the limit $c_s\to 1$,
where we can only constrain
the combination of the two 
parameters, 
\be
{-3.5}\times 10^4<\tilde c_3 (c_s^{-2}-1)<1\times 10^4 \quad \text{(95\%~CL)}\,\,.
\ee
Constraints 
on $c_s$
do not noticeably
improve 
with the aggressive
analysis settings.
This is because 
the $(\fnl^{\rm equil},\fnl^{\rm ortho})$ 
parameter 
space excluded by
these settings
mostly corresponds 
to the 
unphysical 
values 
$c_s^2>1$, $c_s^2<0$,
and hence does not
contribute to 
our final constraints 
on this parameter.\footnote{Notice that flat priors on $(\fnl^{\rm equil},\fnl^{\rm ortho})$ 
do not 
respect the physical conditions $ 0 \leq c_s^2\leq 1$, 
see~Eq.~\eqref{eq:decomposition} and Ref.~\cite{Senatore:2009gt}.}

\textit{Conclusions ---} 
In this \textit{Letter} we have presented the first non-CMB constraints 
on non-local primordial non-Gaussianity, using the BOSS redshift-space clustering data. 
Our nominal constraints 
on the orthogonal shape
and the inflaton sound speed 
are competitive with those 
from the Wilkinson Microwave Anisotropy Probe CMB data~\cite{2013ApJS..208...20B}.
These constraints will
certainly 
improve with the data from upcoming surveys such as \textit{Euclid}~\cite{Laureijs:2011gra}
or DESI~\cite{Aghamousa:2016zmz}. Based on their volume w.r.t.~BOSS, one may expect a reduction of error bars 
by a factor of $3$.
However, the limits from current and future surveys
can be improved even further if more bispectrum 
data is used, e.g.~from new triangles with larger wavenumbers, and from
the angular dependence 
captured by higher order 
harmonics (multipoles) of the redshift-space bispectrum~\cite{Scoccimarro:2015bla}.
We leave including this 
information for future work. 
In addition, it would be interesting
to analyze more complex PNG scenarios, 
and also include information
from the recently measured 
BOSS galaxy trispectrum~\cite{Philcox:2021hbm}.

Finally, it is worth noting 
that the PNG constraints from 
large-scale structure will further improve with futuristic surveys,
such as MegaMapper~\cite{Schlegel:2019eqc}, 
and 21cm/intensity mapping observations,
which will map our Universe
at high redshifts where the 
late-time non-Gaussian clustering background
is weak. 

Our work thus 
serves as a proof of principle,
and
paves the way toward 
systematic analyses 
of PNG with upcoming large-scale 
structure surveys.

\vskip 4pt
\textit{Acknowledgments ---} 
{\small 
We are grateful to Alex Barreira, Enrico Pajer, and Fabian Schmidt for valuable comments on the draft. GC acknowledges support from the Institute for Advanced Study. The work of 
MMI has been supported by NASA through the NASA Hubble Fellowship grant \#HST-HF2-51483.001-A awarded by the Space Telescope Science Institute, which is operated by the Association of Universities for Research in Astronomy, Incorporated, under NASA contract NAS5-26555. OHEP thanks the Simons Foundation for support. Parameter estimates presented in this paper have been obtained with the \texttt{CLASS-PT} 
Boltzmann code \cite{Chudaykin:2020aoj}
(see also \cite{Blas:2011rf}) interfaced with the \texttt{Montepython} MCMC sampler \cite{Audren:2012wb,Brinckmann:2018cvx}. 
The triangle plots are generated with the \texttt{getdist} package\footnote{\href{https://getdist.readthedocs.io/en/latest/}{
\textcolor{blue}{https://getdist.readthedocs.io/en/latest/}}
}~\cite{Lewis:2019xzd}.}

\bibliography{short.bib}

\begin{thebibliography}{85}%
\makeatletter
\providecommand \@ifxundefined [1]{%
 \@ifx{#1\undefined}
}%
\providecommand \@ifnum [1]{%
 \ifnum #1\expandafter \@firstoftwo
 \else \expandafter \@secondoftwo
 \fi
}%
\providecommand \@ifx [1]{%
 \ifx #1\expandafter \@firstoftwo
 \else \expandafter \@secondoftwo
 \fi
}%
\providecommand \natexlab [1]{#1}%
\providecommand \enquote  [1]{``#1''}%
\providecommand \bibnamefont  [1]{#1}%
\providecommand \bibfnamefont [1]{#1}%
\providecommand \citenamefont [1]{#1}%
\providecommand \href@noop [0]{\@secondoftwo}%
\providecommand \href [0]{\begingroup \@sanitize@url \@href}%
\providecommand \@href[1]{\@@startlink{#1}\@@href}%
\providecommand \@@href[1]{\endgroup#1\@@endlink}%
\providecommand \@sanitize@url [0]{\catcode `\\12\catcode `\$12\catcode
  `\&12\catcode `\#12\catcode `\^12\catcode `\_12\catcode `\%12\relax}%
\providecommand \@@startlink[1]{}%
\providecommand \@@endlink[0]{}%
\providecommand \url  [0]{\begingroup\@sanitize@url \@url }%
\providecommand \@url [1]{\endgroup\@href {#1}{\urlprefix }}%
\providecommand \urlprefix  [0]{URL }%
\providecommand \Eprint [0]{\href }%
\providecommand \doibase [0]{http://dx.doi.org/}%
\providecommand \selectlanguage [0]{\@gobble}%
\providecommand \bibinfo  [0]{\@secondoftwo}%
\providecommand \bibfield  [0]{\@secondoftwo}%
\providecommand \translation [1]{[#1]}%
\providecommand \BibitemOpen [0]{}%
\providecommand \bibitemStop [0]{}%
\providecommand \bibitemNoStop [0]{.\EOS\space}%
\providecommand \EOS [0]{\spacefactor3000\relax}%
\providecommand \BibitemShut  [1]{\csname bibitem#1\endcsname}%
\let\auto@bib@innerbib\@empty
\bibitem [{\citenamefont {Linde}\ and\ \citenamefont
  {Mukhanov}(1997)}]{Linde:1996gt}%
  \BibitemOpen
  \bibfield  {author} {\bibinfo {author} {\bibfnamefont {A.~D.}\ \bibnamefont
  {Linde}}\ and\ \bibinfo {author} {\bibfnamefont {V.~F.}\ \bibnamefont
  {Mukhanov}},\ }\href {\doibase 10.1103/PhysRevD.56.R535} {\bibfield
  {journal} {\bibinfo  {journal} {Phys. Rev. D}\ }\textbf {\bibinfo {volume}
  {56}},\ \bibinfo {pages} {R535} (\bibinfo {year} {1997})},\ \Eprint
  {http://arxiv.org/abs/astro-ph/9610219} {arXiv:astro-ph/9610219} \BibitemShut
  {NoStop}%
\bibitem [{\citenamefont {Enqvist}\ and\ \citenamefont
  {Sloth}(2002)}]{Enqvist:2001zp}%
  \BibitemOpen
  \bibfield  {author} {\bibinfo {author} {\bibfnamefont {K.}~\bibnamefont
  {Enqvist}}\ and\ \bibinfo {author} {\bibfnamefont {M.~S.}\ \bibnamefont
  {Sloth}},\ }\href {\doibase 10.1016/S0550-3213(02)00043-3} {\bibfield
  {journal} {\bibinfo  {journal} {Nucl. Phys. B}\ }\textbf {\bibinfo {volume}
  {626}},\ \bibinfo {pages} {395} (\bibinfo {year} {2002})},\ \Eprint
  {http://arxiv.org/abs/hep-ph/0109214} {arXiv:hep-ph/0109214} \BibitemShut
  {NoStop}%
\bibitem [{\citenamefont {Lyth}\ and\ \citenamefont
  {Wands}(2002)}]{Lyth:2001nq}%
  \BibitemOpen
  \bibfield  {author} {\bibinfo {author} {\bibfnamefont {D.~H.}\ \bibnamefont
  {Lyth}}\ and\ \bibinfo {author} {\bibfnamefont {D.}~\bibnamefont {Wands}},\
  }\href {\doibase 10.1016/S0370-2693(01)01366-1} {\bibfield  {journal}
  {\bibinfo  {journal} {Phys. Lett. B}\ }\textbf {\bibinfo {volume} {524}},\
  \bibinfo {pages} {5} (\bibinfo {year} {2002})},\ \Eprint
  {http://arxiv.org/abs/hep-ph/0110002} {arXiv:hep-ph/0110002} \BibitemShut
  {NoStop}%
\bibitem [{\citenamefont {Lyth}\ \emph {et~al.}(2003)\citenamefont {Lyth},
  \citenamefont {Ungarelli},\ and\ \citenamefont {Wands}}]{Lyth:2002my}%
  \BibitemOpen
  \bibfield  {author} {\bibinfo {author} {\bibfnamefont {D.~H.}\ \bibnamefont
  {Lyth}}, \bibinfo {author} {\bibfnamefont {C.}~\bibnamefont {Ungarelli}}, \
  and\ \bibinfo {author} {\bibfnamefont {D.}~\bibnamefont {Wands}},\ }\href
  {\doibase 10.1103/PhysRevD.67.023503} {\bibfield  {journal} {\bibinfo
  {journal} {Phys. Rev. D}\ }\textbf {\bibinfo {volume} {67}},\ \bibinfo
  {pages} {023503} (\bibinfo {year} {2003})},\ \Eprint
  {http://arxiv.org/abs/astro-ph/0208055} {arXiv:astro-ph/0208055} \BibitemShut
  {NoStop}%
\bibitem [{\citenamefont {Dvali}\ \emph
  {et~al.}(2004{\natexlab{a}})\citenamefont {Dvali}, \citenamefont {Gruzinov},\
  and\ \citenamefont {Zaldarriaga}}]{Dvali:2003em}%
  \BibitemOpen
  \bibfield  {author} {\bibinfo {author} {\bibfnamefont {G.}~\bibnamefont
  {Dvali}}, \bibinfo {author} {\bibfnamefont {A.}~\bibnamefont {Gruzinov}}, \
  and\ \bibinfo {author} {\bibfnamefont {M.}~\bibnamefont {Zaldarriaga}},\
  }\href {\doibase 10.1103/PhysRevD.69.023505} {\bibfield  {journal} {\bibinfo
  {journal} {Phys. Rev. D}\ }\textbf {\bibinfo {volume} {69}},\ \bibinfo
  {pages} {023505} (\bibinfo {year} {2004}{\natexlab{a}})},\ \Eprint
  {http://arxiv.org/abs/astro-ph/0303591} {arXiv:astro-ph/0303591} \BibitemShut
  {NoStop}%
\bibitem [{\citenamefont {Dvali}\ \emph
  {et~al.}(2004{\natexlab{b}})\citenamefont {Dvali}, \citenamefont {Gruzinov},\
  and\ \citenamefont {Zaldarriaga}}]{Dvali:2003ar}%
  \BibitemOpen
  \bibfield  {author} {\bibinfo {author} {\bibfnamefont {G.}~\bibnamefont
  {Dvali}}, \bibinfo {author} {\bibfnamefont {A.}~\bibnamefont {Gruzinov}}, \
  and\ \bibinfo {author} {\bibfnamefont {M.}~\bibnamefont {Zaldarriaga}},\
  }\href {\doibase 10.1103/PhysRevD.69.083505} {\bibfield  {journal} {\bibinfo
  {journal} {Phys. Rev. D}\ }\textbf {\bibinfo {volume} {69}},\ \bibinfo
  {pages} {083505} (\bibinfo {year} {2004}{\natexlab{b}})},\ \Eprint
  {http://arxiv.org/abs/astro-ph/0305548} {arXiv:astro-ph/0305548} \BibitemShut
  {NoStop}%
\bibitem [{\citenamefont {Zaldarriaga}(2004)}]{Zaldarriaga:2003my}%
  \BibitemOpen
  \bibfield  {author} {\bibinfo {author} {\bibfnamefont {M.}~\bibnamefont
  {Zaldarriaga}},\ }\href {\doibase 10.1103/PhysRevD.69.043508} {\bibfield
  {journal} {\bibinfo  {journal} {Phys. Rev. D}\ }\textbf {\bibinfo {volume}
  {69}},\ \bibinfo {pages} {043508} (\bibinfo {year} {2004})},\ \Eprint
  {http://arxiv.org/abs/astro-ph/0306006} {arXiv:astro-ph/0306006} \BibitemShut
  {NoStop}%
\bibitem [{\citenamefont {Byrnes}\ and\ \citenamefont
  {Choi}(2010)}]{Byrnes:2010em}%
  \BibitemOpen
  \bibfield  {author} {\bibinfo {author} {\bibfnamefont {C.~T.}\ \bibnamefont
  {Byrnes}}\ and\ \bibinfo {author} {\bibfnamefont {K.-Y.}\ \bibnamefont
  {Choi}},\ }\href {\doibase 10.1155/2010/724525} {\bibfield  {journal}
  {\bibinfo  {journal} {Adv. Astron.}\ }\textbf {\bibinfo {volume} {2010}},\
  \bibinfo {pages} {724525} (\bibinfo {year} {2010})},\ \Eprint
  {http://arxiv.org/abs/1002.3110} {arXiv:1002.3110 [astro-ph.CO]} \BibitemShut
  {NoStop}%
\bibitem [{\citenamefont {Senatore}\ and\ \citenamefont
  {Zaldarriaga}(2012)}]{Senatore:2010wk}%
  \BibitemOpen
  \bibfield  {author} {\bibinfo {author} {\bibfnamefont {L.}~\bibnamefont
  {Senatore}}\ and\ \bibinfo {author} {\bibfnamefont {M.}~\bibnamefont
  {Zaldarriaga}},\ }\href {\doibase 10.1007/JHEP04(2012)024} {\bibfield
  {journal} {\bibinfo  {journal} {JHEP}\ }\textbf {\bibinfo {volume} {04}},\
  \bibinfo {pages} {024} (\bibinfo {year} {2012})},\ \Eprint
  {http://arxiv.org/abs/1009.2093} {arXiv:1009.2093 [hep-th]} \BibitemShut
  {NoStop}%
\bibitem [{\citenamefont {Lehners}(2010)}]{Lehners:2010fy}%
  \BibitemOpen
  \bibfield  {author} {\bibinfo {author} {\bibfnamefont {J.-L.}\ \bibnamefont
  {Lehners}},\ }\href {\doibase 10.1155/2010/903907} {\bibfield  {journal}
  {\bibinfo  {journal} {Adv. Astron.}\ }\textbf {\bibinfo {volume} {2010}},\
  \bibinfo {pages} {903907} (\bibinfo {year} {2010})},\ \Eprint
  {http://arxiv.org/abs/1001.3125} {arXiv:1001.3125 [hep-th]} \BibitemShut
  {NoStop}%
\bibitem [{\citenamefont {Akrami}\ \emph {et~al.}(2020)\citenamefont {Akrami}
  \emph {et~al.}}]{Planck:2019kim}%
  \BibitemOpen
  \bibfield  {author} {\bibinfo {author} {\bibfnamefont {Y.}~\bibnamefont
  {Akrami}} \emph {et~al.} (\bibinfo {collaboration} {Planck}),\ }\href
  {\doibase 10.1051/0004-6361/201935891} {\bibfield  {journal} {\bibinfo
  {journal} {Astron. Astrophys.}\ }\textbf {\bibinfo {volume} {641}},\ \bibinfo
  {pages} {A9} (\bibinfo {year} {2020})},\ \Eprint
  {http://arxiv.org/abs/1905.05697} {arXiv:1905.05697 [astro-ph.CO]}
  \BibitemShut {NoStop}%
\bibitem [{\citenamefont {Arkani-Hamed}\ \emph {et~al.}(2004)\citenamefont
  {Arkani-Hamed}, \citenamefont {Creminelli}, \citenamefont {Mukohyama},\ and\
  \citenamefont {Zaldarriaga}}]{Arkani-Hamed:2003juy}%
  \BibitemOpen
  \bibfield  {author} {\bibinfo {author} {\bibfnamefont {N.}~\bibnamefont
  {Arkani-Hamed}}, \bibinfo {author} {\bibfnamefont {P.}~\bibnamefont
  {Creminelli}}, \bibinfo {author} {\bibfnamefont {S.}~\bibnamefont
  {Mukohyama}}, \ and\ \bibinfo {author} {\bibfnamefont {M.}~\bibnamefont
  {Zaldarriaga}},\ }\href {\doibase 10.1088/1475-7516/2004/04/001} {\bibfield
  {journal} {\bibinfo  {journal} {JCAP}\ }\textbf {\bibinfo {volume} {04}},\
  \bibinfo {pages} {001} (\bibinfo {year} {2004})},\ \Eprint
  {http://arxiv.org/abs/hep-th/0312100} {arXiv:hep-th/0312100} \BibitemShut
  {NoStop}%
\bibitem [{\citenamefont {Alishahiha}\ \emph {et~al.}(2004)\citenamefont
  {Alishahiha}, \citenamefont {Silverstein},\ and\ \citenamefont
  {Tong}}]{Alishahiha:2004eh}%
  \BibitemOpen
  \bibfield  {author} {\bibinfo {author} {\bibfnamefont {M.}~\bibnamefont
  {Alishahiha}}, \bibinfo {author} {\bibfnamefont {E.}~\bibnamefont
  {Silverstein}}, \ and\ \bibinfo {author} {\bibfnamefont {D.}~\bibnamefont
  {Tong}},\ }\href {\doibase 10.1103/PhysRevD.70.123505} {\bibfield  {journal}
  {\bibinfo  {journal} {Phys. Rev. D}\ }\textbf {\bibinfo {volume} {70}},\
  \bibinfo {pages} {123505} (\bibinfo {year} {2004})},\ \Eprint
  {http://arxiv.org/abs/hep-th/0404084} {arXiv:hep-th/0404084} \BibitemShut
  {NoStop}%
\bibitem [{\citenamefont {Senatore}(2005)}]{Senatore:2004rj}%
  \BibitemOpen
  \bibfield  {author} {\bibinfo {author} {\bibfnamefont {L.}~\bibnamefont
  {Senatore}},\ }\href {\doibase 10.1103/PhysRevD.71.043512} {\bibfield
  {journal} {\bibinfo  {journal} {Phys. Rev. D}\ }\textbf {\bibinfo {volume}
  {71}},\ \bibinfo {pages} {043512} (\bibinfo {year} {2005})},\ \Eprint
  {http://arxiv.org/abs/astro-ph/0406187} {arXiv:astro-ph/0406187} \BibitemShut
  {NoStop}%
\bibitem [{\citenamefont {Chen}\ \emph {et~al.}(2007)\citenamefont {Chen},
  \citenamefont {Huang}, \citenamefont {Kachru},\ and\ \citenamefont
  {Shiu}}]{Chen:2006nt}%
  \BibitemOpen
  \bibfield  {author} {\bibinfo {author} {\bibfnamefont {X.}~\bibnamefont
  {Chen}}, \bibinfo {author} {\bibfnamefont {M.-x.}\ \bibnamefont {Huang}},
  \bibinfo {author} {\bibfnamefont {S.}~\bibnamefont {Kachru}}, \ and\ \bibinfo
  {author} {\bibfnamefont {G.}~\bibnamefont {Shiu}},\ }\href {\doibase
  10.1088/1475-7516/2007/01/002} {\bibfield  {journal} {\bibinfo  {journal}
  {JCAP}\ }\textbf {\bibinfo {volume} {01}},\ \bibinfo {pages} {002} (\bibinfo
  {year} {2007})},\ \Eprint {http://arxiv.org/abs/hep-th/0605045}
  {arXiv:hep-th/0605045} \BibitemShut {NoStop}%
\bibitem [{\citenamefont {Creminelli}\ \emph {et~al.}(2006)\citenamefont
  {Creminelli}, \citenamefont {Luty}, \citenamefont {Nicolis},\ and\
  \citenamefont {Senatore}}]{Creminelli:2006xe}%
  \BibitemOpen
  \bibfield  {author} {\bibinfo {author} {\bibfnamefont {P.}~\bibnamefont
  {Creminelli}}, \bibinfo {author} {\bibfnamefont {M.~A.}\ \bibnamefont
  {Luty}}, \bibinfo {author} {\bibfnamefont {A.}~\bibnamefont {Nicolis}}, \
  and\ \bibinfo {author} {\bibfnamefont {L.}~\bibnamefont {Senatore}},\ }\href
  {\doibase 10.1088/1126-6708/2006/12/080} {\bibfield  {journal} {\bibinfo
  {journal} {JHEP}\ }\textbf {\bibinfo {volume} {12}},\ \bibinfo {pages} {080}
  (\bibinfo {year} {2006})},\ \Eprint {http://arxiv.org/abs/hep-th/0606090}
  {arXiv:hep-th/0606090} \BibitemShut {NoStop}%
\bibitem [{\citenamefont {Cheung}\ \emph
  {et~al.}(2008{\natexlab{a}})\citenamefont {Cheung}, \citenamefont
  {Creminelli}, \citenamefont {Fitzpatrick}, \citenamefont {Kaplan},\ and\
  \citenamefont {Senatore}}]{Cheung:2007st}%
  \BibitemOpen
  \bibfield  {author} {\bibinfo {author} {\bibfnamefont {C.}~\bibnamefont
  {Cheung}}, \bibinfo {author} {\bibfnamefont {P.}~\bibnamefont {Creminelli}},
  \bibinfo {author} {\bibfnamefont {A.~L.}\ \bibnamefont {Fitzpatrick}},
  \bibinfo {author} {\bibfnamefont {J.}~\bibnamefont {Kaplan}}, \ and\ \bibinfo
  {author} {\bibfnamefont {L.}~\bibnamefont {Senatore}},\ }\href {\doibase
  10.1088/1126-6708/2008/03/014} {\bibfield  {journal} {\bibinfo  {journal}
  {JHEP}\ }\textbf {\bibinfo {volume} {03}},\ \bibinfo {pages} {014} (\bibinfo
  {year} {2008}{\natexlab{a}})},\ \Eprint {http://arxiv.org/abs/0709.0293}
  {arXiv:0709.0293 [hep-th]} \BibitemShut {NoStop}%
\bibitem [{\citenamefont {Cheung}\ \emph
  {et~al.}(2008{\natexlab{b}})\citenamefont {Cheung}, \citenamefont
  {Fitzpatrick}, \citenamefont {Kaplan},\ and\ \citenamefont
  {Senatore}}]{Cheung:2007sv}%
  \BibitemOpen
  \bibfield  {author} {\bibinfo {author} {\bibfnamefont {C.}~\bibnamefont
  {Cheung}}, \bibinfo {author} {\bibfnamefont {A.~L.}\ \bibnamefont
  {Fitzpatrick}}, \bibinfo {author} {\bibfnamefont {J.}~\bibnamefont {Kaplan}},
  \ and\ \bibinfo {author} {\bibfnamefont {L.}~\bibnamefont {Senatore}},\
  }\href {\doibase 10.1088/1475-7516/2008/02/021} {\bibfield  {journal}
  {\bibinfo  {journal} {JCAP}\ }\textbf {\bibinfo {volume} {02}},\ \bibinfo
  {pages} {021} (\bibinfo {year} {2008}{\natexlab{b}})},\ \Eprint
  {http://arxiv.org/abs/0709.0295} {arXiv:0709.0295 [hep-th]} \BibitemShut
  {NoStop}%
\bibitem [{\citenamefont {Senatore}\ \emph {et~al.}(2010)\citenamefont
  {Senatore}, \citenamefont {Smith},\ and\ \citenamefont
  {Zaldarriaga}}]{Senatore:2009gt}%
  \BibitemOpen
  \bibfield  {author} {\bibinfo {author} {\bibfnamefont {L.}~\bibnamefont
  {Senatore}}, \bibinfo {author} {\bibfnamefont {K.~M.}\ \bibnamefont {Smith}},
  \ and\ \bibinfo {author} {\bibfnamefont {M.}~\bibnamefont {Zaldarriaga}},\
  }\href {\doibase 10.1088/1475-7516/2010/01/028} {\bibfield  {journal}
  {\bibinfo  {journal} {JCAP}\ }\textbf {\bibinfo {volume} {01}},\ \bibinfo
  {pages} {028} (\bibinfo {year} {2010})},\ \Eprint
  {http://arxiv.org/abs/0905.3746} {arXiv:0905.3746 [astro-ph.CO]} \BibitemShut
  {NoStop}%
\bibitem [{\citenamefont {Baumann}\ \emph {et~al.}(2016)\citenamefont
  {Baumann}, \citenamefont {Green}, \citenamefont {Lee},\ and\ \citenamefont
  {Porto}}]{Baumann:2015nta}%
  \BibitemOpen
  \bibfield  {author} {\bibinfo {author} {\bibfnamefont {D.}~\bibnamefont
  {Baumann}}, \bibinfo {author} {\bibfnamefont {D.}~\bibnamefont {Green}},
  \bibinfo {author} {\bibfnamefont {H.}~\bibnamefont {Lee}}, \ and\ \bibinfo
  {author} {\bibfnamefont {R.~A.}\ \bibnamefont {Porto}},\ }\href {\doibase
  10.1103/PhysRevD.93.023523} {\bibfield  {journal} {\bibinfo  {journal} {Phys.
  Rev. D}\ }\textbf {\bibinfo {volume} {93}},\ \bibinfo {pages} {023523}
  (\bibinfo {year} {2016})},\ \Eprint {http://arxiv.org/abs/1502.07304}
  {arXiv:1502.07304 [hep-th]} \BibitemShut {NoStop}%
\bibitem [{\citenamefont {Green}\ and\ \citenamefont
  {Pajer}(2020)}]{Green:2020ebl}%
  \BibitemOpen
  \bibfield  {author} {\bibinfo {author} {\bibfnamefont {D.}~\bibnamefont
  {Green}}\ and\ \bibinfo {author} {\bibfnamefont {E.}~\bibnamefont {Pajer}},\
  }\href {\doibase 10.1088/1475-7516/2020/09/032} {\bibfield  {journal}
  {\bibinfo  {journal} {JCAP}\ }\textbf {\bibinfo {volume} {09}},\ \bibinfo
  {pages} {032} (\bibinfo {year} {2020})},\ \Eprint
  {http://arxiv.org/abs/2004.09587} {arXiv:2004.09587 [hep-th]} \BibitemShut
  {NoStop}%
\bibitem [{\citenamefont {{Bennett}}\ \emph {et~al.}(2013)\citenamefont
  {{Bennett}}, \citenamefont {{Larson}}, \citenamefont {{Weiland}},
  \citenamefont {{Jarosik}}, \citenamefont {{Hinshaw}}, \citenamefont
  {{Odegard}}, \citenamefont {{Smith}}, \citenamefont {{Hill}}, \citenamefont
  {{Gold}}, \citenamefont {{Halpern}}, \citenamefont {{Komatsu}}, \citenamefont
  {{Nolta}}, \citenamefont {{Page}}, \citenamefont {{Spergel}}, \citenamefont
  {{Wollack}}, \citenamefont {{Dunkley}}, \citenamefont {{Kogut}},
  \citenamefont {{Limon}}, \citenamefont {{Meyer}}, \citenamefont {{Tucker}},\
  and\ \citenamefont {{Wright}}}]{2013ApJS..208...20B}%
  \BibitemOpen
  \bibfield  {author} {\bibinfo {author} {\bibfnamefont {C.~L.}\ \bibnamefont
  {{Bennett}}}, \bibinfo {author} {\bibfnamefont {D.}~\bibnamefont {{Larson}}},
  \bibinfo {author} {\bibfnamefont {J.~L.}\ \bibnamefont {{Weiland}}}, \bibinfo
  {author} {\bibfnamefont {N.}~\bibnamefont {{Jarosik}}}, \bibinfo {author}
  {\bibfnamefont {G.}~\bibnamefont {{Hinshaw}}}, \bibinfo {author}
  {\bibfnamefont {N.}~\bibnamefont {{Odegard}}}, \bibinfo {author}
  {\bibfnamefont {K.~M.}\ \bibnamefont {{Smith}}}, \bibinfo {author}
  {\bibfnamefont {R.~S.}\ \bibnamefont {{Hill}}}, \bibinfo {author}
  {\bibfnamefont {B.}~\bibnamefont {{Gold}}}, \bibinfo {author} {\bibfnamefont
  {M.}~\bibnamefont {{Halpern}}}, \bibinfo {author} {\bibfnamefont
  {E.}~\bibnamefont {{Komatsu}}}, \bibinfo {author} {\bibfnamefont {M.~R.}\
  \bibnamefont {{Nolta}}}, \bibinfo {author} {\bibfnamefont {L.}~\bibnamefont
  {{Page}}}, \bibinfo {author} {\bibfnamefont {D.~N.}\ \bibnamefont
  {{Spergel}}}, \bibinfo {author} {\bibfnamefont {E.}~\bibnamefont
  {{Wollack}}}, \bibinfo {author} {\bibfnamefont {J.}~\bibnamefont
  {{Dunkley}}}, \bibinfo {author} {\bibfnamefont {A.}~\bibnamefont {{Kogut}}},
  \bibinfo {author} {\bibfnamefont {M.}~\bibnamefont {{Limon}}}, \bibinfo
  {author} {\bibfnamefont {S.~S.}\ \bibnamefont {{Meyer}}}, \bibinfo {author}
  {\bibfnamefont {G.~S.}\ \bibnamefont {{Tucker}}}, \ and\ \bibinfo {author}
  {\bibfnamefont {E.~L.}\ \bibnamefont {{Wright}}},\ }\href {\doibase
  10.1088/0067-0049/208/2/20} {\bibfield  {journal} {\bibinfo  {journal}
  {"Astrophys. J."}\ }\textbf {\bibinfo {volume} {208}},\ \bibinfo {eid} {20}
  (\bibinfo {year} {2013})},\ \Eprint {http://arxiv.org/abs/1212.5225}
  {arXiv:1212.5225 [astro-ph.CO]} \BibitemShut {NoStop}%
\bibitem [{\citenamefont {Leistedt}\ \emph {et~al.}(2014)\citenamefont
  {Leistedt}, \citenamefont {Peiris},\ and\ \citenamefont
  {Roth}}]{Leistedt:2014zqa}%
  \BibitemOpen
  \bibfield  {author} {\bibinfo {author} {\bibfnamefont {B.}~\bibnamefont
  {Leistedt}}, \bibinfo {author} {\bibfnamefont {H.~V.}\ \bibnamefont
  {Peiris}}, \ and\ \bibinfo {author} {\bibfnamefont {N.}~\bibnamefont
  {Roth}},\ }\href {\doibase 10.1103/PhysRevLett.113.221301} {\bibfield
  {journal} {\bibinfo  {journal} {Phys. Rev. Lett.}\ }\textbf {\bibinfo
  {volume} {113}},\ \bibinfo {pages} {221301} (\bibinfo {year} {2014})},\
  \Eprint {http://arxiv.org/abs/1405.4315} {arXiv:1405.4315 [astro-ph.CO]}
  \BibitemShut {NoStop}%
\bibitem [{\citenamefont {Castorina}\ \emph {et~al.}(2019)\citenamefont
  {Castorina} \emph {et~al.}}]{Castorina:2019wmr}%
  \BibitemOpen
  \bibfield  {author} {\bibinfo {author} {\bibfnamefont {E.}~\bibnamefont
  {Castorina}} \emph {et~al.},\ }\href {\doibase 10.1088/1475-7516/2019/09/010}
  {\bibfield  {journal} {\bibinfo  {journal} {JCAP}\ }\textbf {\bibinfo
  {volume} {09}},\ \bibinfo {pages} {010} (\bibinfo {year} {2019})},\ \Eprint
  {http://arxiv.org/abs/1904.08859} {arXiv:1904.08859 [astro-ph.CO]}
  \BibitemShut {NoStop}%
\bibitem [{\citenamefont {Mueller}\ \emph {et~al.}(2021)\citenamefont {Mueller}
  \emph {et~al.}}]{Mueller:2021tqa}%
  \BibitemOpen
  \bibfield  {author} {\bibinfo {author} {\bibfnamefont {E.-M.}\ \bibnamefont
  {Mueller}} \emph {et~al.},\ }\href@noop {} {\  (\bibinfo {year} {2021})},\
  \Eprint {http://arxiv.org/abs/2106.13725} {arXiv:2106.13725 [astro-ph.CO]}
  \BibitemShut {NoStop}%
\bibitem [{\citenamefont {Ivanov}\ \emph
  {et~al.}(2021{\natexlab{a}})\citenamefont {Ivanov}, \citenamefont {Philcox},
  \citenamefont {Nishimichi}, \citenamefont {Simonovi\'c}, \citenamefont
  {Takada},\ and\ \citenamefont {Zaldarriaga}}]{Ivanov:2021kcd}%
  \BibitemOpen
  \bibfield  {author} {\bibinfo {author} {\bibfnamefont {M.~M.}\ \bibnamefont
  {Ivanov}}, \bibinfo {author} {\bibfnamefont {O.~H.~E.}\ \bibnamefont
  {Philcox}}, \bibinfo {author} {\bibfnamefont {T.}~\bibnamefont {Nishimichi}},
  \bibinfo {author} {\bibfnamefont {M.}~\bibnamefont {Simonovi\'c}}, \bibinfo
  {author} {\bibfnamefont {M.}~\bibnamefont {Takada}}, \ and\ \bibinfo {author}
  {\bibfnamefont {M.}~\bibnamefont {Zaldarriaga}},\ }\href@noop {} {\
  (\bibinfo {year} {2021}{\natexlab{a}})},\ \Eprint
  {http://arxiv.org/abs/2110.10161} {arXiv:2110.10161 [astro-ph.CO]}
  \BibitemShut {NoStop}%
\bibitem [{\citenamefont {Philcox}\ and\ \citenamefont
  {Ivanov}(2021)}]{Philcox:2021kcw}%
  \BibitemOpen
  \bibfield  {author} {\bibinfo {author} {\bibfnamefont {O.~H.~E.}\
  \bibnamefont {Philcox}}\ and\ \bibinfo {author} {\bibfnamefont {M.~M.}\
  \bibnamefont {Ivanov}},\ }\href@noop {} {\  (\bibinfo {year} {2021})},\
  \Eprint {http://arxiv.org/abs/2112.04515} {arXiv:2112.04515 [astro-ph.CO]}
  \BibitemShut {NoStop}%
\bibitem [{\citenamefont {Scoccimarro}(2000)}]{Scoccimarro:2000sn}%
  \BibitemOpen
  \bibfield  {author} {\bibinfo {author} {\bibfnamefont {R.}~\bibnamefont
  {Scoccimarro}},\ }\href {\doibase 10.1086/317248} {\bibfield  {journal}
  {\bibinfo  {journal} {Astrophys. J.}\ }\textbf {\bibinfo {volume} {544}},\
  \bibinfo {pages} {597} (\bibinfo {year} {2000})},\ \Eprint
  {http://arxiv.org/abs/astro-ph/0004086} {arXiv:astro-ph/0004086} \BibitemShut
  {NoStop}%
\bibitem [{\citenamefont {Sefusatti}\ \emph {et~al.}(2006)\citenamefont
  {Sefusatti}, \citenamefont {Crocce}, \citenamefont {Pueblas},\ and\
  \citenamefont {Scoccimarro}}]{Sefusatti:2006pa}%
  \BibitemOpen
  \bibfield  {author} {\bibinfo {author} {\bibfnamefont {E.}~\bibnamefont
  {Sefusatti}}, \bibinfo {author} {\bibfnamefont {M.}~\bibnamefont {Crocce}},
  \bibinfo {author} {\bibfnamefont {S.}~\bibnamefont {Pueblas}}, \ and\
  \bibinfo {author} {\bibfnamefont {R.}~\bibnamefont {Scoccimarro}},\ }\href
  {\doibase 10.1103/PhysRevD.74.023522} {\bibfield  {journal} {\bibinfo
  {journal} {Phys. Rev.}\ }\textbf {\bibinfo {volume} {D74}},\ \bibinfo {pages}
  {023522} (\bibinfo {year} {2006})},\ \Eprint
  {http://arxiv.org/abs/astro-ph/0604505} {arXiv:astro-ph/0604505 [astro-ph]}
  \BibitemShut {NoStop}%
\bibitem [{\citenamefont {Baldauf}\ \emph
  {et~al.}(2015{\natexlab{a}})\citenamefont {Baldauf}, \citenamefont
  {Mercolli}, \citenamefont {Mirbabayi},\ and\ \citenamefont
  {Pajer}}]{Baldauf:2014qfa}%
  \BibitemOpen
  \bibfield  {author} {\bibinfo {author} {\bibfnamefont {T.}~\bibnamefont
  {Baldauf}}, \bibinfo {author} {\bibfnamefont {L.}~\bibnamefont {Mercolli}},
  \bibinfo {author} {\bibfnamefont {M.}~\bibnamefont {Mirbabayi}}, \ and\
  \bibinfo {author} {\bibfnamefont {E.}~\bibnamefont {Pajer}},\ }\href
  {\doibase 10.1088/1475-7516/2015/05/007} {\bibfield  {journal} {\bibinfo
  {journal} {JCAP}\ }\textbf {\bibinfo {volume} {1505}},\ \bibinfo {pages}
  {007} (\bibinfo {year} {2015}{\natexlab{a}})},\ \Eprint
  {http://arxiv.org/abs/1406.4135} {arXiv:1406.4135 [astro-ph.CO]} \BibitemShut
  {NoStop}%
\bibitem [{\citenamefont {Angulo}\ \emph
  {et~al.}(2015{\natexlab{a}})\citenamefont {Angulo}, \citenamefont {Foreman},
  \citenamefont {Schmittfull},\ and\ \citenamefont
  {Senatore}}]{Angulo:2014tfa}%
  \BibitemOpen
  \bibfield  {author} {\bibinfo {author} {\bibfnamefont {R.~E.}\ \bibnamefont
  {Angulo}}, \bibinfo {author} {\bibfnamefont {S.}~\bibnamefont {Foreman}},
  \bibinfo {author} {\bibfnamefont {M.}~\bibnamefont {Schmittfull}}, \ and\
  \bibinfo {author} {\bibfnamefont {L.}~\bibnamefont {Senatore}},\ }\href
  {\doibase 10.1088/1475-7516/2015/10/039} {\bibfield  {journal} {\bibinfo
  {journal} {JCAP}\ }\textbf {\bibinfo {volume} {1510}},\ \bibinfo {pages}
  {039} (\bibinfo {year} {2015}{\natexlab{a}})},\ \Eprint
  {http://arxiv.org/abs/1406.4143} {arXiv:1406.4143 [astro-ph.CO]} \BibitemShut
  {NoStop}%
\bibitem [{\citenamefont {Eggemeier}\ \emph {et~al.}(2018)\citenamefont
  {Eggemeier}, \citenamefont {Scoccimarro},\ and\ \citenamefont
  {Smith}}]{Eggemeier:2018qae}%
  \BibitemOpen
  \bibfield  {author} {\bibinfo {author} {\bibfnamefont {A.}~\bibnamefont
  {Eggemeier}}, \bibinfo {author} {\bibfnamefont {R.}~\bibnamefont
  {Scoccimarro}}, \ and\ \bibinfo {author} {\bibfnamefont {R.~E.}\ \bibnamefont
  {Smith}},\ }\href@noop {} {\  (\bibinfo {year} {2018})},\ \Eprint
  {http://arxiv.org/abs/1812.03208} {arXiv:1812.03208 [astro-ph.CO]}
  \BibitemShut {NoStop}%
\bibitem [{\citenamefont {Ivanov}\ \emph
  {et~al.}(2020{\natexlab{a}})\citenamefont {Ivanov}, \citenamefont
  {Simonovi\'c},\ and\ \citenamefont {Zaldarriaga}}]{Ivanov:2019pdj}%
  \BibitemOpen
  \bibfield  {author} {\bibinfo {author} {\bibfnamefont {M.~M.}\ \bibnamefont
  {Ivanov}}, \bibinfo {author} {\bibfnamefont {M.}~\bibnamefont {Simonovi\'c}},
  \ and\ \bibinfo {author} {\bibfnamefont {M.}~\bibnamefont {Zaldarriaga}},\
  }\href {\doibase 10.1088/1475-7516/2020/05/042} {\bibfield  {journal}
  {\bibinfo  {journal} {JCAP}\ }\textbf {\bibinfo {volume} {05}},\ \bibinfo
  {pages} {042} (\bibinfo {year} {2020}{\natexlab{a}})},\ \Eprint
  {http://arxiv.org/abs/1909.05277} {arXiv:1909.05277 [astro-ph.CO]}
  \BibitemShut {NoStop}%
\bibitem [{\citenamefont {D'Amico}\ \emph {et~al.}(2019)\citenamefont
  {D'Amico}, \citenamefont {Gleyzes}, \citenamefont {Kokron}, \citenamefont
  {Markovic}, \citenamefont {Senatore}, \citenamefont {Zhang}, \citenamefont
  {Beutler},\ and\ \citenamefont {Gil-Marín}}]{DAmico:2019fhj}%
  \BibitemOpen
  \bibfield  {author} {\bibinfo {author} {\bibfnamefont {G.}~\bibnamefont
  {D'Amico}}, \bibinfo {author} {\bibfnamefont {J.}~\bibnamefont {Gleyzes}},
  \bibinfo {author} {\bibfnamefont {N.}~\bibnamefont {Kokron}}, \bibinfo
  {author} {\bibfnamefont {D.}~\bibnamefont {Markovic}}, \bibinfo {author}
  {\bibfnamefont {L.}~\bibnamefont {Senatore}}, \bibinfo {author}
  {\bibfnamefont {P.}~\bibnamefont {Zhang}}, \bibinfo {author} {\bibfnamefont
  {F.}~\bibnamefont {Beutler}}, \ and\ \bibinfo {author} {\bibfnamefont
  {H.}~\bibnamefont {Gil-Marín}},\ }\href@noop {} {\  (\bibinfo {year}
  {2019})},\ \Eprint {http://arxiv.org/abs/1909.05271} {arXiv:1909.05271
  [astro-ph.CO]} \BibitemShut {NoStop}%
\bibitem [{\citenamefont {Oddo}\ \emph {et~al.}(2020)\citenamefont {Oddo},
  \citenamefont {Sefusatti}, \citenamefont {Porciani}, \citenamefont {Monaco},\
  and\ \citenamefont {S\'anchez}}]{Oddo:2019run}%
  \BibitemOpen
  \bibfield  {author} {\bibinfo {author} {\bibfnamefont {A.}~\bibnamefont
  {Oddo}}, \bibinfo {author} {\bibfnamefont {E.}~\bibnamefont {Sefusatti}},
  \bibinfo {author} {\bibfnamefont {C.}~\bibnamefont {Porciani}}, \bibinfo
  {author} {\bibfnamefont {P.}~\bibnamefont {Monaco}}, \ and\ \bibinfo {author}
  {\bibfnamefont {A.~G.}\ \bibnamefont {S\'anchez}},\ }\href {\doibase
  10.1088/1475-7516/2020/03/056} {\bibfield  {journal} {\bibinfo  {journal}
  {JCAP}\ }\textbf {\bibinfo {volume} {03}},\ \bibinfo {pages} {056} (\bibinfo
  {year} {2020})},\ \Eprint {http://arxiv.org/abs/1908.01774} {arXiv:1908.01774
  [astro-ph.CO]} \BibitemShut {NoStop}%
\bibitem [{\citenamefont {Eggemeier}\ \emph {et~al.}(2021)\citenamefont
  {Eggemeier}, \citenamefont {Scoccimarro}, \citenamefont {Smith},
  \citenamefont {Crocce}, \citenamefont {Pezzotta},\ and\ \citenamefont
  {S\'anchez}}]{Eggemeier:2021cam}%
  \BibitemOpen
  \bibfield  {author} {\bibinfo {author} {\bibfnamefont {A.}~\bibnamefont
  {Eggemeier}}, \bibinfo {author} {\bibfnamefont {R.}~\bibnamefont
  {Scoccimarro}}, \bibinfo {author} {\bibfnamefont {R.~E.}\ \bibnamefont
  {Smith}}, \bibinfo {author} {\bibfnamefont {M.}~\bibnamefont {Crocce}},
  \bibinfo {author} {\bibfnamefont {A.}~\bibnamefont {Pezzotta}}, \ and\
  \bibinfo {author} {\bibfnamefont {A.~G.}\ \bibnamefont {S\'anchez}},\
  }\href@noop {} {\  (\bibinfo {year} {2021})},\ \Eprint
  {http://arxiv.org/abs/2102.06902} {arXiv:2102.06902 [astro-ph.CO]}
  \BibitemShut {NoStop}%
\bibitem [{\citenamefont {Alkhanishvili}\ \emph {et~al.}(2021)\citenamefont
  {Alkhanishvili}, \citenamefont {Porciani}, \citenamefont {Sefusatti},
  \citenamefont {Biagetti}, \citenamefont {Lazanu}, \citenamefont {Oddo},\ and\
  \citenamefont {Yankelevich}}]{Alkhanishvili:2021pvy}%
  \BibitemOpen
  \bibfield  {author} {\bibinfo {author} {\bibfnamefont {D.}~\bibnamefont
  {Alkhanishvili}}, \bibinfo {author} {\bibfnamefont {C.}~\bibnamefont
  {Porciani}}, \bibinfo {author} {\bibfnamefont {E.}~\bibnamefont {Sefusatti}},
  \bibinfo {author} {\bibfnamefont {M.}~\bibnamefont {Biagetti}}, \bibinfo
  {author} {\bibfnamefont {A.}~\bibnamefont {Lazanu}}, \bibinfo {author}
  {\bibfnamefont {A.}~\bibnamefont {Oddo}}, \ and\ \bibinfo {author}
  {\bibfnamefont {V.}~\bibnamefont {Yankelevich}},\ }\href@noop {} {\
  (\bibinfo {year} {2021})},\ \Eprint {http://arxiv.org/abs/2107.08054}
  {arXiv:2107.08054 [astro-ph.CO]} \BibitemShut {NoStop}%
\bibitem [{\citenamefont {Oddo}\ \emph {et~al.}(2021)\citenamefont {Oddo},
  \citenamefont {Rizzo}, \citenamefont {Sefusatti}, \citenamefont {Porciani},\
  and\ \citenamefont {Monaco}}]{Oddo:2021iwq}%
  \BibitemOpen
  \bibfield  {author} {\bibinfo {author} {\bibfnamefont {A.}~\bibnamefont
  {Oddo}}, \bibinfo {author} {\bibfnamefont {F.}~\bibnamefont {Rizzo}},
  \bibinfo {author} {\bibfnamefont {E.}~\bibnamefont {Sefusatti}}, \bibinfo
  {author} {\bibfnamefont {C.}~\bibnamefont {Porciani}}, \ and\ \bibinfo
  {author} {\bibfnamefont {P.}~\bibnamefont {Monaco}},\ }\href {\doibase
  10.1088/1475-7516/2021/11/038} {\bibfield  {journal} {\bibinfo  {journal}
  {JCAP}\ }\textbf {\bibinfo {volume} {11}},\ \bibinfo {pages} {038} (\bibinfo
  {year} {2021})},\ \Eprint {http://arxiv.org/abs/2108.03204} {arXiv:2108.03204
  [astro-ph.CO]} \BibitemShut {NoStop}%
\bibitem [{\citenamefont {Chen}\ \emph {et~al.}(2021)\citenamefont {Chen},
  \citenamefont {Vlah},\ and\ \citenamefont {White}}]{Chen:2021wdi}%
  \BibitemOpen
  \bibfield  {author} {\bibinfo {author} {\bibfnamefont {S.-F.}\ \bibnamefont
  {Chen}}, \bibinfo {author} {\bibfnamefont {Z.}~\bibnamefont {Vlah}}, \ and\
  \bibinfo {author} {\bibfnamefont {M.}~\bibnamefont {White}},\ }\href@noop {}
  {\  (\bibinfo {year} {2021})},\ \Eprint {http://arxiv.org/abs/2110.05530}
  {arXiv:2110.05530 [astro-ph.CO]} \BibitemShut {NoStop}%
\bibitem [{\citenamefont {Baldauf}\ \emph {et~al.}(2021)\citenamefont
  {Baldauf}, \citenamefont {Garny}, \citenamefont {Taule},\ and\ \citenamefont
  {Steele}}]{Baldauf:2021zlt}%
  \BibitemOpen
  \bibfield  {author} {\bibinfo {author} {\bibfnamefont {T.}~\bibnamefont
  {Baldauf}}, \bibinfo {author} {\bibfnamefont {M.}~\bibnamefont {Garny}},
  \bibinfo {author} {\bibfnamefont {P.}~\bibnamefont {Taule}}, \ and\ \bibinfo
  {author} {\bibfnamefont {T.}~\bibnamefont {Steele}},\ }\href {\doibase
  10.1103/PhysRevD.104.123551} {\bibfield  {journal} {\bibinfo  {journal}
  {Phys. Rev. D}\ }\textbf {\bibinfo {volume} {104}},\ \bibinfo {pages}
  {123551} (\bibinfo {year} {2021})},\ \Eprint
  {http://arxiv.org/abs/2110.13930} {arXiv:2110.13930 [astro-ph.CO]}
  \BibitemShut {NoStop}%
\bibitem [{\citenamefont {Philcox}(2021{\natexlab{a}})}]{Philcox:2020vbm}%
  \BibitemOpen
  \bibfield  {author} {\bibinfo {author} {\bibfnamefont {O.~H.~E.}\
  \bibnamefont {Philcox}},\ }\href {\doibase 10.1103/PhysRevD.103.103504}
  {\bibfield  {journal} {\bibinfo  {journal} {Phys. Rev. D}\ }\textbf {\bibinfo
  {volume} {103}},\ \bibinfo {pages} {103504} (\bibinfo {year}
  {2021}{\natexlab{a}})},\ \Eprint {http://arxiv.org/abs/2012.09389}
  {arXiv:2012.09389 [astro-ph.CO]} \BibitemShut {NoStop}%
\bibitem [{\citenamefont {Philcox}(2021{\natexlab{b}})}]{Philcox:2021ukg}%
  \BibitemOpen
  \bibfield  {author} {\bibinfo {author} {\bibfnamefont {O.~H.~E.}\
  \bibnamefont {Philcox}},\ }\href {\doibase 10.1103/PhysRevD.104.123529}
  {\bibfield  {journal} {\bibinfo  {journal} {Phys. Rev. D}\ }\textbf {\bibinfo
  {volume} {104}},\ \bibinfo {pages} {123529} (\bibinfo {year}
  {2021}{\natexlab{b}})},\ \Eprint {http://arxiv.org/abs/2107.06287}
  {arXiv:2107.06287 [astro-ph.CO]} \BibitemShut {NoStop}%
\bibitem [{\citenamefont {Aghanim}\ \emph {et~al.}(2018)\citenamefont {Aghanim}
  \emph {et~al.}}]{Aghanim:2018eyx}%
  \BibitemOpen
  \bibfield  {author} {\bibinfo {author} {\bibfnamefont {N.}~\bibnamefont
  {Aghanim}} \emph {et~al.} (\bibinfo {collaboration} {Planck}),\ }\href@noop
  {} {\  (\bibinfo {year} {2018})},\ \Eprint {http://arxiv.org/abs/1807.06209}
  {arXiv:1807.06209 [astro-ph.CO]} \BibitemShut {NoStop}%
\bibitem [{\citenamefont {Babich}\ \emph {et~al.}(2004)\citenamefont {Babich},
  \citenamefont {Creminelli},\ and\ \citenamefont
  {Zaldarriaga}}]{Babich:2004gb}%
  \BibitemOpen
  \bibfield  {author} {\bibinfo {author} {\bibfnamefont {D.}~\bibnamefont
  {Babich}}, \bibinfo {author} {\bibfnamefont {P.}~\bibnamefont {Creminelli}},
  \ and\ \bibinfo {author} {\bibfnamefont {M.}~\bibnamefont {Zaldarriaga}},\
  }\href {\doibase 10.1088/1475-7516/2004/08/009} {\bibfield  {journal}
  {\bibinfo  {journal} {JCAP}\ }\textbf {\bibinfo {volume} {08}},\ \bibinfo
  {pages} {009} (\bibinfo {year} {2004})},\ \Eprint
  {http://arxiv.org/abs/astro-ph/0405356} {arXiv:astro-ph/0405356} \BibitemShut
  {NoStop}%
\bibitem [{\citenamefont {Baumann}\ \emph {et~al.}(2012)\citenamefont
  {Baumann}, \citenamefont {Nicolis}, \citenamefont {Senatore},\ and\
  \citenamefont {Zaldarriaga}}]{Baumann:2010tm}%
  \BibitemOpen
  \bibfield  {author} {\bibinfo {author} {\bibfnamefont {D.}~\bibnamefont
  {Baumann}}, \bibinfo {author} {\bibfnamefont {A.}~\bibnamefont {Nicolis}},
  \bibinfo {author} {\bibfnamefont {L.}~\bibnamefont {Senatore}}, \ and\
  \bibinfo {author} {\bibfnamefont {M.}~\bibnamefont {Zaldarriaga}},\ }\href
  {\doibase 10.1088/1475-7516/2012/07/051} {\bibfield  {journal} {\bibinfo
  {journal} {JCAP}\ }\textbf {\bibinfo {volume} {1207}},\ \bibinfo {pages}
  {051} (\bibinfo {year} {2012})},\ \Eprint {http://arxiv.org/abs/1004.2488}
  {arXiv:1004.2488 [astro-ph.CO]} \BibitemShut {NoStop}%
\bibitem [{\citenamefont {Carrasco}\ \emph {et~al.}(2012)\citenamefont
  {Carrasco}, \citenamefont {Hertzberg},\ and\ \citenamefont
  {Senatore}}]{Carrasco:2012cv}%
  \BibitemOpen
  \bibfield  {author} {\bibinfo {author} {\bibfnamefont {J.~J.~M.}\
  \bibnamefont {Carrasco}}, \bibinfo {author} {\bibfnamefont {M.~P.}\
  \bibnamefont {Hertzberg}}, \ and\ \bibinfo {author} {\bibfnamefont
  {L.}~\bibnamefont {Senatore}},\ }\href {\doibase 10.1007/JHEP09(2012)082}
  {\bibfield  {journal} {\bibinfo  {journal} {JHEP}\ }\textbf {\bibinfo
  {volume} {09}},\ \bibinfo {pages} {082} (\bibinfo {year} {2012})},\ \Eprint
  {http://arxiv.org/abs/1206.2926} {arXiv:1206.2926 [astro-ph.CO]} \BibitemShut
  {NoStop}%
\bibitem [{\citenamefont {Desjacques}\ \emph {et~al.}(2018)\citenamefont
  {Desjacques}, \citenamefont {Jeong},\ and\ \citenamefont
  {Schmidt}}]{Desjacques:2016bnm}%
  \BibitemOpen
  \bibfield  {author} {\bibinfo {author} {\bibfnamefont {V.}~\bibnamefont
  {Desjacques}}, \bibinfo {author} {\bibfnamefont {D.}~\bibnamefont {Jeong}}, \
  and\ \bibinfo {author} {\bibfnamefont {F.}~\bibnamefont {Schmidt}},\ }\href
  {\doibase 10.1016/j.physrep.2017.12.002} {\bibfield  {journal} {\bibinfo
  {journal} {Phys. Rept.}\ }\textbf {\bibinfo {volume} {733}},\ \bibinfo
  {pages} {1} (\bibinfo {year} {2018})},\ \Eprint
  {http://arxiv.org/abs/1611.09787} {arXiv:1611.09787 [astro-ph.CO]}
  \BibitemShut {NoStop}%
\bibitem [{\citenamefont {Nishimichi}\ \emph {et~al.}(2020)\citenamefont
  {Nishimichi}, \citenamefont {D'Amico}, \citenamefont {Ivanov}, \citenamefont
  {Senatore}, \citenamefont {Simonovi\'c}, \citenamefont {Takada},
  \citenamefont {Zaldarriaga},\ and\ \citenamefont
  {Zhang}}]{Nishimichi:2020tvu}%
  \BibitemOpen
  \bibfield  {author} {\bibinfo {author} {\bibfnamefont {T.}~\bibnamefont
  {Nishimichi}}, \bibinfo {author} {\bibfnamefont {G.}~\bibnamefont {D'Amico}},
  \bibinfo {author} {\bibfnamefont {M.~M.}\ \bibnamefont {Ivanov}}, \bibinfo
  {author} {\bibfnamefont {L.}~\bibnamefont {Senatore}}, \bibinfo {author}
  {\bibfnamefont {M.}~\bibnamefont {Simonovi\'c}}, \bibinfo {author}
  {\bibfnamefont {M.}~\bibnamefont {Takada}}, \bibinfo {author} {\bibfnamefont
  {M.}~\bibnamefont {Zaldarriaga}}, \ and\ \bibinfo {author} {\bibfnamefont
  {P.}~\bibnamefont {Zhang}},\ }\href {\doibase 10.1103/PhysRevD.102.123541}
  {\bibfield  {journal} {\bibinfo  {journal} {Phys. Rev. D}\ }\textbf {\bibinfo
  {volume} {102}},\ \bibinfo {pages} {123541} (\bibinfo {year} {2020})},\
  \Eprint {http://arxiv.org/abs/2003.08277} {arXiv:2003.08277 [astro-ph.CO]}
  \BibitemShut {NoStop}%
\bibitem [{\citenamefont {Sefusatti}\ and\ \citenamefont
  {Komatsu}(2007)}]{Sefusatti:2007ih}%
  \BibitemOpen
  \bibfield  {author} {\bibinfo {author} {\bibfnamefont {E.}~\bibnamefont
  {Sefusatti}}\ and\ \bibinfo {author} {\bibfnamefont {E.}~\bibnamefont
  {Komatsu}},\ }\href {\doibase 10.1103/PhysRevD.76.083004} {\bibfield
  {journal} {\bibinfo  {journal} {Phys. Rev. D}\ }\textbf {\bibinfo {volume}
  {76}},\ \bibinfo {pages} {083004} (\bibinfo {year} {2007})},\ \Eprint
  {http://arxiv.org/abs/0705.0343} {arXiv:0705.0343 [astro-ph]} \BibitemShut
  {NoStop}%
\bibitem [{\citenamefont {Sefusatti}(2009)}]{Sefusatti:2009qh}%
  \BibitemOpen
  \bibfield  {author} {\bibinfo {author} {\bibfnamefont {E.}~\bibnamefont
  {Sefusatti}},\ }\href {\doibase 10.1103/PhysRevD.80.123002} {\bibfield
  {journal} {\bibinfo  {journal} {Phys. Rev. D}\ }\textbf {\bibinfo {volume}
  {80}},\ \bibinfo {pages} {123002} (\bibinfo {year} {2009})},\ \Eprint
  {http://arxiv.org/abs/0905.0717} {arXiv:0905.0717 [astro-ph.CO]} \BibitemShut
  {NoStop}%
\bibitem [{\citenamefont {Chudaykin}\ \emph {et~al.}(2020)\citenamefont
  {Chudaykin}, \citenamefont {Ivanov}, \citenamefont {Philcox},\ and\
  \citenamefont {Simonovi\'c}}]{Chudaykin:2020aoj}%
  \BibitemOpen
  \bibfield  {author} {\bibinfo {author} {\bibfnamefont {A.}~\bibnamefont
  {Chudaykin}}, \bibinfo {author} {\bibfnamefont {M.~M.}\ \bibnamefont
  {Ivanov}}, \bibinfo {author} {\bibfnamefont {O.~H.~E.}\ \bibnamefont
  {Philcox}}, \ and\ \bibinfo {author} {\bibfnamefont {M.}~\bibnamefont
  {Simonovi\'c}},\ }\href {\doibase 10.1103/PhysRevD.102.063533} {\bibfield
  {journal} {\bibinfo  {journal} {Phys. Rev. D}\ }\textbf {\bibinfo {volume}
  {102}},\ \bibinfo {pages} {063533} (\bibinfo {year} {2020})},\ \Eprint
  {http://arxiv.org/abs/2004.10607} {arXiv:2004.10607 [astro-ph.CO]}
  \BibitemShut {NoStop}%
\bibitem [{\citenamefont {Taruya}\ \emph {et~al.}(2008)\citenamefont {Taruya},
  \citenamefont {Koyama},\ and\ \citenamefont {Matsubara}}]{Taruya:2008pg}%
  \BibitemOpen
  \bibfield  {author} {\bibinfo {author} {\bibfnamefont {A.}~\bibnamefont
  {Taruya}}, \bibinfo {author} {\bibfnamefont {K.}~\bibnamefont {Koyama}}, \
  and\ \bibinfo {author} {\bibfnamefont {T.}~\bibnamefont {Matsubara}},\ }\href
  {\doibase 10.1103/PhysRevD.78.123534} {\bibfield  {journal} {\bibinfo
  {journal} {Phys. Rev. D}\ }\textbf {\bibinfo {volume} {78}},\ \bibinfo
  {pages} {123534} (\bibinfo {year} {2008})},\ \Eprint
  {http://arxiv.org/abs/0808.4085} {arXiv:0808.4085 [astro-ph]} \BibitemShut
  {NoStop}%
\bibitem [{\citenamefont {Assassi}\ \emph
  {et~al.}(2015{\natexlab{a}})\citenamefont {Assassi}, \citenamefont {Baumann},
  \citenamefont {Pajer}, \citenamefont {Welling},\ and\ \citenamefont {van~der
  Woude}}]{Assassi:2015jqa}%
  \BibitemOpen
  \bibfield  {author} {\bibinfo {author} {\bibfnamefont {V.}~\bibnamefont
  {Assassi}}, \bibinfo {author} {\bibfnamefont {D.}~\bibnamefont {Baumann}},
  \bibinfo {author} {\bibfnamefont {E.}~\bibnamefont {Pajer}}, \bibinfo
  {author} {\bibfnamefont {Y.}~\bibnamefont {Welling}}, \ and\ \bibinfo
  {author} {\bibfnamefont {D.}~\bibnamefont {van~der Woude}},\ }\href {\doibase
  10.1088/1475-7516/2015/11/024} {\bibfield  {journal} {\bibinfo  {journal}
  {JCAP}\ }\textbf {\bibinfo {volume} {11}},\ \bibinfo {pages} {024} (\bibinfo
  {year} {2015}{\natexlab{a}})},\ \Eprint {http://arxiv.org/abs/1505.06668}
  {arXiv:1505.06668 [astro-ph.CO]} \BibitemShut {NoStop}%
\bibitem [{\citenamefont {Moradinezhad~Dizgah}\ \emph
  {et~al.}(2021)\citenamefont {Moradinezhad~Dizgah}, \citenamefont {Biagetti},
  \citenamefont {Sefusatti}, \citenamefont {Desjacques},\ and\ \citenamefont
  {Nore\~na}}]{MoradinezhadDizgah:2020whw}%
  \BibitemOpen
  \bibfield  {author} {\bibinfo {author} {\bibfnamefont {A.}~\bibnamefont
  {Moradinezhad~Dizgah}}, \bibinfo {author} {\bibfnamefont {M.}~\bibnamefont
  {Biagetti}}, \bibinfo {author} {\bibfnamefont {E.}~\bibnamefont {Sefusatti}},
  \bibinfo {author} {\bibfnamefont {V.}~\bibnamefont {Desjacques}}, \ and\
  \bibinfo {author} {\bibfnamefont {J.}~\bibnamefont {Nore\~na}},\ }\href
  {\doibase 10.1088/1475-7516/2021/05/015} {\bibfield  {journal} {\bibinfo
  {journal} {JCAP}\ }\textbf {\bibinfo {volume} {05}},\ \bibinfo {pages} {015}
  (\bibinfo {year} {2021})},\ \Eprint {http://arxiv.org/abs/2010.14523}
  {arXiv:2010.14523 [astro-ph.CO]} \BibitemShut {NoStop}%
\bibitem [{\citenamefont {Assassi}\ \emph
  {et~al.}(2015{\natexlab{b}})\citenamefont {Assassi}, \citenamefont
  {Baumann},\ and\ \citenamefont {Schmidt}}]{Assassi:2015fma}%
  \BibitemOpen
  \bibfield  {author} {\bibinfo {author} {\bibfnamefont {V.}~\bibnamefont
  {Assassi}}, \bibinfo {author} {\bibfnamefont {D.}~\bibnamefont {Baumann}}, \
  and\ \bibinfo {author} {\bibfnamefont {F.}~\bibnamefont {Schmidt}},\ }\href
  {\doibase 10.1088/1475-7516/2015/12/043} {\bibfield  {journal} {\bibinfo
  {journal} {JCAP}\ }\textbf {\bibinfo {volume} {12}},\ \bibinfo {pages} {043}
  (\bibinfo {year} {2015}{\natexlab{b}})},\ \Eprint
  {http://arxiv.org/abs/1510.03723} {arXiv:1510.03723 [astro-ph.CO]}
  \BibitemShut {NoStop}%
\bibitem [{\citenamefont {Angulo}\ \emph
  {et~al.}(2015{\natexlab{b}})\citenamefont {Angulo}, \citenamefont {Fasiello},
  \citenamefont {Senatore},\ and\ \citenamefont {Vlah}}]{Angulo:2015eqa}%
  \BibitemOpen
  \bibfield  {author} {\bibinfo {author} {\bibfnamefont {R.}~\bibnamefont
  {Angulo}}, \bibinfo {author} {\bibfnamefont {M.}~\bibnamefont {Fasiello}},
  \bibinfo {author} {\bibfnamefont {L.}~\bibnamefont {Senatore}}, \ and\
  \bibinfo {author} {\bibfnamefont {Z.}~\bibnamefont {Vlah}},\ }\href {\doibase
  10.1088/1475-7516/2015/09/029, 10.1088/1475-7516/2015/9/029} {\bibfield
  {journal} {\bibinfo  {journal} {JCAP}\ }\textbf {\bibinfo {volume} {1509}},\
  \bibinfo {pages} {029} (\bibinfo {year} {2015}{\natexlab{b}})},\ \Eprint
  {http://arxiv.org/abs/1503.08826} {arXiv:1503.08826 [astro-ph.CO]}
  \BibitemShut {NoStop}%
\bibitem [{\citenamefont {Pajer}\ and\ \citenamefont
  {Zaldarriaga}(2013)}]{Pajer:2013jj}%
  \BibitemOpen
  \bibfield  {author} {\bibinfo {author} {\bibfnamefont {E.}~\bibnamefont
  {Pajer}}\ and\ \bibinfo {author} {\bibfnamefont {M.}~\bibnamefont
  {Zaldarriaga}},\ }\href {\doibase 10.1088/1475-7516/2013/08/037} {\bibfield
  {journal} {\bibinfo  {journal} {JCAP}\ }\textbf {\bibinfo {volume} {08}},\
  \bibinfo {pages} {037} (\bibinfo {year} {2013})},\ \Eprint
  {http://arxiv.org/abs/1301.7182} {arXiv:1301.7182 [astro-ph.CO]} \BibitemShut
  {NoStop}%
\bibitem [{\citenamefont {Ivanov}\ \emph
  {et~al.}(2020{\natexlab{b}})\citenamefont {Ivanov}, \citenamefont
  {Simonovi\'c},\ and\ \citenamefont {Zaldarriaga}}]{Ivanov:2019hqk}%
  \BibitemOpen
  \bibfield  {author} {\bibinfo {author} {\bibfnamefont {M.~M.}\ \bibnamefont
  {Ivanov}}, \bibinfo {author} {\bibfnamefont {M.}~\bibnamefont {Simonovi\'c}},
  \ and\ \bibinfo {author} {\bibfnamefont {M.}~\bibnamefont {Zaldarriaga}},\
  }\href {\doibase 10.1103/PhysRevD.101.083504} {\bibfield  {journal} {\bibinfo
   {journal} {Phys. Rev. D}\ }\textbf {\bibinfo {volume} {101}},\ \bibinfo
  {pages} {083504} (\bibinfo {year} {2020}{\natexlab{b}})},\ \Eprint
  {http://arxiv.org/abs/1912.08208} {arXiv:1912.08208 [astro-ph.CO]}
  \BibitemShut {NoStop}%
\bibitem [{\citenamefont {Senatore}\ and\ \citenamefont
  {Zaldarriaga}(2015)}]{Senatore:2014via}%
  \BibitemOpen
  \bibfield  {author} {\bibinfo {author} {\bibfnamefont {L.}~\bibnamefont
  {Senatore}}\ and\ \bibinfo {author} {\bibfnamefont {M.}~\bibnamefont
  {Zaldarriaga}},\ }\href {\doibase 10.1088/1475-7516/2015/02/013} {\bibfield
  {journal} {\bibinfo  {journal} {JCAP}\ }\textbf {\bibinfo {volume} {1502}},\
  \bibinfo {pages} {013} (\bibinfo {year} {2015})},\ \Eprint
  {http://arxiv.org/abs/1404.5954} {arXiv:1404.5954 [astro-ph.CO]} \BibitemShut
  {NoStop}%
\bibitem [{\citenamefont {Baldauf}\ \emph
  {et~al.}(2015{\natexlab{b}})\citenamefont {Baldauf}, \citenamefont
  {Mirbabayi}, \citenamefont {Simonović},\ and\ \citenamefont
  {Zaldarriaga}}]{Baldauf:2015xfa}%
  \BibitemOpen
  \bibfield  {author} {\bibinfo {author} {\bibfnamefont {T.}~\bibnamefont
  {Baldauf}}, \bibinfo {author} {\bibfnamefont {M.}~\bibnamefont {Mirbabayi}},
  \bibinfo {author} {\bibfnamefont {M.}~\bibnamefont {Simonović}}, \ and\
  \bibinfo {author} {\bibfnamefont {M.}~\bibnamefont {Zaldarriaga}},\ }\href
  {\doibase 10.1103/PhysRevD.92.043514} {\bibfield  {journal} {\bibinfo
  {journal} {Phys. Rev.}\ }\textbf {\bibinfo {volume} {D92}},\ \bibinfo {pages}
  {043514} (\bibinfo {year} {2015}{\natexlab{b}})},\ \Eprint
  {http://arxiv.org/abs/1504.04366} {arXiv:1504.04366 [astro-ph.CO]}
  \BibitemShut {NoStop}%
\bibitem [{\citenamefont {Blas}\ \emph
  {et~al.}(2016{\natexlab{a}})\citenamefont {Blas}, \citenamefont {Garny},
  \citenamefont {Ivanov},\ and\ \citenamefont {Sibiryakov}}]{Blas:2015qsi}%
  \BibitemOpen
  \bibfield  {author} {\bibinfo {author} {\bibfnamefont {D.}~\bibnamefont
  {Blas}}, \bibinfo {author} {\bibfnamefont {M.}~\bibnamefont {Garny}},
  \bibinfo {author} {\bibfnamefont {M.~M.}\ \bibnamefont {Ivanov}}, \ and\
  \bibinfo {author} {\bibfnamefont {S.}~\bibnamefont {Sibiryakov}},\ }\href
  {\doibase 10.1088/1475-7516/2016/07/052} {\bibfield  {journal} {\bibinfo
  {journal} {JCAP}\ }\textbf {\bibinfo {volume} {1607}},\ \bibinfo {pages}
  {052} (\bibinfo {year} {2016}{\natexlab{a}})},\ \Eprint
  {http://arxiv.org/abs/1512.05807} {arXiv:1512.05807 [astro-ph.CO]}
  \BibitemShut {NoStop}%
\bibitem [{\citenamefont {Blas}\ \emph
  {et~al.}(2016{\natexlab{b}})\citenamefont {Blas}, \citenamefont {Garny},
  \citenamefont {Ivanov},\ and\ \citenamefont {Sibiryakov}}]{Blas:2016sfa}%
  \BibitemOpen
  \bibfield  {author} {\bibinfo {author} {\bibfnamefont {D.}~\bibnamefont
  {Blas}}, \bibinfo {author} {\bibfnamefont {M.}~\bibnamefont {Garny}},
  \bibinfo {author} {\bibfnamefont {M.~M.}\ \bibnamefont {Ivanov}}, \ and\
  \bibinfo {author} {\bibfnamefont {S.}~\bibnamefont {Sibiryakov}},\ }\href
  {\doibase 10.1088/1475-7516/2016/07/028} {\bibfield  {journal} {\bibinfo
  {journal} {JCAP}\ }\textbf {\bibinfo {volume} {1607}},\ \bibinfo {pages}
  {028} (\bibinfo {year} {2016}{\natexlab{b}})},\ \Eprint
  {http://arxiv.org/abs/1605.02149} {arXiv:1605.02149 [astro-ph.CO]}
  \BibitemShut {NoStop}%
\bibitem [{\citenamefont {Ivanov}\ and\ \citenamefont
  {Sibiryakov}(2018)}]{Ivanov:2018gjr}%
  \BibitemOpen
  \bibfield  {author} {\bibinfo {author} {\bibfnamefont {M.~M.}\ \bibnamefont
  {Ivanov}}\ and\ \bibinfo {author} {\bibfnamefont {S.}~\bibnamefont
  {Sibiryakov}},\ }\href {\doibase 10.1088/1475-7516/2018/07/053} {\bibfield
  {journal} {\bibinfo  {journal} {JCAP}\ }\textbf {\bibinfo {volume} {1807}},\
  \bibinfo {pages} {053} (\bibinfo {year} {2018})},\ \Eprint
  {http://arxiv.org/abs/1804.05080} {arXiv:1804.05080 [astro-ph.CO]}
  \BibitemShut {NoStop}%
\bibitem [{\citenamefont {Vasudevan}\ \emph {et~al.}(2019)\citenamefont
  {Vasudevan}, \citenamefont {Ivanov}, \citenamefont {Sibiryakov},\ and\
  \citenamefont {Lesgourgues}}]{Vasudevan:2019ewf}%
  \BibitemOpen
  \bibfield  {author} {\bibinfo {author} {\bibfnamefont {A.}~\bibnamefont
  {Vasudevan}}, \bibinfo {author} {\bibfnamefont {M.~M.}\ \bibnamefont
  {Ivanov}}, \bibinfo {author} {\bibfnamefont {S.}~\bibnamefont {Sibiryakov}},
  \ and\ \bibinfo {author} {\bibfnamefont {J.}~\bibnamefont {Lesgourgues}},\
  }\href {\doibase 10.1088/1475-7516/2019/09/037} {\bibfield  {journal}
  {\bibinfo  {journal} {JCAP}\ }\textbf {\bibinfo {volume} {09}},\ \bibinfo
  {pages} {037} (\bibinfo {year} {2019})},\ \Eprint
  {http://arxiv.org/abs/1906.08697} {arXiv:1906.08697 [astro-ph.CO]}
  \BibitemShut {NoStop}%
\bibitem [{\citenamefont {Alcock}\ and\ \citenamefont
  {Paczynski}(1979)}]{Alcock:1979mp}%
  \BibitemOpen
  \bibfield  {author} {\bibinfo {author} {\bibfnamefont {C.}~\bibnamefont
  {Alcock}}\ and\ \bibinfo {author} {\bibfnamefont {B.}~\bibnamefont
  {Paczynski}},\ }\href {\doibase 10.1038/281358a0} {\bibfield  {journal}
  {\bibinfo  {journal} {Nature}\ }\textbf {\bibinfo {volume} {281}},\ \bibinfo
  {pages} {358} (\bibinfo {year} {1979})}\BibitemShut {NoStop}%
\bibitem [{\citenamefont {Alam}\ \emph {et~al.}(2017)\citenamefont {Alam} \emph
  {et~al.}}]{Alam:2016hwk}%
  \BibitemOpen
  \bibfield  {author} {\bibinfo {author} {\bibfnamefont {S.}~\bibnamefont
  {Alam}} \emph {et~al.} (\bibinfo {collaboration} {BOSS}),\ }\href {\doibase
  10.1093/mnras/stx721} {\bibfield  {journal} {\bibinfo  {journal} {Mon. Not.
  Roy. Astron. Soc.}\ }\textbf {\bibinfo {volume} {470}},\ \bibinfo {pages}
  {2617} (\bibinfo {year} {2017})},\ \Eprint {http://arxiv.org/abs/1607.03155}
  {arXiv:1607.03155 [astro-ph.CO]} \BibitemShut {NoStop}%
\bibitem [{\citenamefont {Ivanov}\ \emph
  {et~al.}(2021{\natexlab{b}})\citenamefont {Ivanov}, \citenamefont {Philcox},
  \citenamefont {Simonovi\'c}, \citenamefont {Zaldarriaga}, \citenamefont
  {Nishimichi},\ and\ \citenamefont {Takada}}]{Ivanov:2021haa}%
  \BibitemOpen
  \bibfield  {author} {\bibinfo {author} {\bibfnamefont {M.~M.}\ \bibnamefont
  {Ivanov}}, \bibinfo {author} {\bibfnamefont {O.~H.~E.}\ \bibnamefont
  {Philcox}}, \bibinfo {author} {\bibfnamefont {M.}~\bibnamefont
  {Simonovi\'c}}, \bibinfo {author} {\bibfnamefont {M.}~\bibnamefont
  {Zaldarriaga}}, \bibinfo {author} {\bibfnamefont {T.}~\bibnamefont
  {Nishimichi}}, \ and\ \bibinfo {author} {\bibfnamefont {M.}~\bibnamefont
  {Takada}},\ }\href@noop {} {\  (\bibinfo {year} {2021}{\natexlab{b}})},\
  \Eprint {http://arxiv.org/abs/2110.00006} {arXiv:2110.00006 [astro-ph.CO]}
  \BibitemShut {NoStop}%
\bibitem [{\citenamefont {Philcox}\ \emph {et~al.}(2020)\citenamefont
  {Philcox}, \citenamefont {Ivanov}, \citenamefont {Simonovi\'c},\ and\
  \citenamefont {Zaldarriaga}}]{Philcox:2020vvt}%
  \BibitemOpen
  \bibfield  {author} {\bibinfo {author} {\bibfnamefont {O.~H.~E.}\
  \bibnamefont {Philcox}}, \bibinfo {author} {\bibfnamefont {M.~M.}\
  \bibnamefont {Ivanov}}, \bibinfo {author} {\bibfnamefont {M.}~\bibnamefont
  {Simonovi\'c}}, \ and\ \bibinfo {author} {\bibfnamefont {M.}~\bibnamefont
  {Zaldarriaga}},\ }\href {\doibase 10.1088/1475-7516/2020/05/032} {\bibfield
  {journal} {\bibinfo  {journal} {JCAP}\ }\textbf {\bibinfo {volume} {05}},\
  \bibinfo {pages} {032} (\bibinfo {year} {2020})},\ \Eprint
  {http://arxiv.org/abs/2002.04035} {arXiv:2002.04035 [astro-ph.CO]}
  \BibitemShut {NoStop}%
\bibitem [{\citenamefont {Kitaura}\ \emph {et~al.}(2016)\citenamefont {Kitaura}
  \emph {et~al.}}]{Kitaura:2015uqa}%
  \BibitemOpen
  \bibfield  {author} {\bibinfo {author} {\bibfnamefont {F.-S.}\ \bibnamefont
  {Kitaura}} \emph {et~al.},\ }\href {\doibase 10.1093/mnras/stv2826}
  {\bibfield  {journal} {\bibinfo  {journal} {Mon. Not. Roy. Astron. Soc.}\
  }\textbf {\bibinfo {volume} {456}},\ \bibinfo {pages} {4156} (\bibinfo {year}
  {2016})},\ \Eprint {http://arxiv.org/abs/1509.06400} {arXiv:1509.06400
  [astro-ph.CO]} \BibitemShut {NoStop}%
\bibitem [{\citenamefont {Simonović}\ \emph {et~al.}(2018)\citenamefont
  {Simonović}, \citenamefont {Baldauf}, \citenamefont {Zaldarriaga},
  \citenamefont {Carrasco},\ and\ \citenamefont
  {Kollmeier}}]{Simonovic:2017mhp}%
  \BibitemOpen
  \bibfield  {author} {\bibinfo {author} {\bibfnamefont {M.}~\bibnamefont
  {Simonović}}, \bibinfo {author} {\bibfnamefont {T.}~\bibnamefont {Baldauf}},
  \bibinfo {author} {\bibfnamefont {M.}~\bibnamefont {Zaldarriaga}}, \bibinfo
  {author} {\bibfnamefont {J.~J.}\ \bibnamefont {Carrasco}}, \ and\ \bibinfo
  {author} {\bibfnamefont {J.~A.}\ \bibnamefont {Kollmeier}},\ }\href {\doibase
  10.1088/1475-7516/2018/04/030} {\bibfield  {journal} {\bibinfo  {journal}
  {JCAP}\ }\textbf {\bibinfo {volume} {1804}},\ \bibinfo {pages} {030}
  (\bibinfo {year} {2018})},\ \Eprint {http://arxiv.org/abs/1708.08130}
  {arXiv:1708.08130 [astro-ph.CO]} \BibitemShut {NoStop}%
\bibitem [{\citenamefont {Cooke}\ \emph {et~al.}(2018)\citenamefont {Cooke},
  \citenamefont {Pettini},\ and\ \citenamefont {Steidel}}]{Cooke:2017cwo}%
  \BibitemOpen
  \bibfield  {author} {\bibinfo {author} {\bibfnamefont {R.~J.}\ \bibnamefont
  {Cooke}}, \bibinfo {author} {\bibfnamefont {M.}~\bibnamefont {Pettini}}, \
  and\ \bibinfo {author} {\bibfnamefont {C.~C.}\ \bibnamefont {Steidel}},\
  }\href {\doibase 10.3847/1538-4357/aaab53} {\bibfield  {journal} {\bibinfo
  {journal} {Astrophys. J.}\ }\textbf {\bibinfo {volume} {855}},\ \bibinfo
  {pages} {102} (\bibinfo {year} {2018})},\ \Eprint
  {http://arxiv.org/abs/1710.11129} {arXiv:1710.11129 [astro-ph.CO]}
  \BibitemShut {NoStop}%
\bibitem [{\citenamefont {Schmidt}\ and\ \citenamefont
  {Kamionkowski}(2010)}]{Schmidt:2010gw}%
  \BibitemOpen
  \bibfield  {author} {\bibinfo {author} {\bibfnamefont {F.}~\bibnamefont
  {Schmidt}}\ and\ \bibinfo {author} {\bibfnamefont {M.}~\bibnamefont
  {Kamionkowski}},\ }\href {\doibase 10.1103/PhysRevD.82.103002} {\bibfield
  {journal} {\bibinfo  {journal} {Phys. Rev. D}\ }\textbf {\bibinfo {volume}
  {82}},\ \bibinfo {pages} {103002} (\bibinfo {year} {2010})},\ \Eprint
  {http://arxiv.org/abs/1008.0638} {arXiv:1008.0638 [astro-ph.CO]} \BibitemShut
  {NoStop}%
\bibitem [{\citenamefont {Barreira}\ \emph {et~al.}(2021)\citenamefont
  {Barreira}, \citenamefont {Lazeyras},\ and\ \citenamefont
  {Schmidt}}]{Barreira:2021ukk}%
  \BibitemOpen
  \bibfield  {author} {\bibinfo {author} {\bibfnamefont {A.}~\bibnamefont
  {Barreira}}, \bibinfo {author} {\bibfnamefont {T.}~\bibnamefont {Lazeyras}},
  \ and\ \bibinfo {author} {\bibfnamefont {F.}~\bibnamefont {Schmidt}},\
  }\href@noop {} {\  (\bibinfo {year} {2021})},\ \Eprint
  {http://arxiv.org/abs/2105.02876} {arXiv:2105.02876 [astro-ph.CO]}
  \BibitemShut {NoStop}%
\bibitem [{\citenamefont {Lazeyras}\ \emph {et~al.}(2021)\citenamefont
  {Lazeyras}, \citenamefont {Barreira},\ and\ \citenamefont
  {Schmidt}}]{Lazeyras:2021dar}%
  \BibitemOpen
  \bibfield  {author} {\bibinfo {author} {\bibfnamefont {T.}~\bibnamefont
  {Lazeyras}}, \bibinfo {author} {\bibfnamefont {A.}~\bibnamefont {Barreira}},
  \ and\ \bibinfo {author} {\bibfnamefont {F.}~\bibnamefont {Schmidt}},\ }\href
  {\doibase 10.1088/1475-7516/2021/10/063} {\bibfield  {journal} {\bibinfo
  {journal} {JCAP}\ }\textbf {\bibinfo {volume} {10}},\ \bibinfo {pages} {063}
  (\bibinfo {year} {2021})},\ \Eprint {http://arxiv.org/abs/2106.14713}
  {arXiv:2106.14713 [astro-ph.CO]} \BibitemShut {NoStop}%
\bibitem [{\citenamefont {Laureijs}\ \emph {et~al.}(2011)\citenamefont
  {Laureijs} \emph {et~al.}}]{Laureijs:2011gra}%
  \BibitemOpen
  \bibfield  {author} {\bibinfo {author} {\bibfnamefont {R.}~\bibnamefont
  {Laureijs}} \emph {et~al.} (\bibinfo {collaboration} {EUCLID}),\ }\href@noop
  {} {\  (\bibinfo {year} {2011})},\ \Eprint {http://arxiv.org/abs/1110.3193}
  {arXiv:1110.3193 [astro-ph.CO]} \BibitemShut {NoStop}%
\bibitem [{\citenamefont {Aghamousa}\ \emph {et~al.}(2016)\citenamefont
  {Aghamousa} \emph {et~al.}}]{Aghamousa:2016zmz}%
  \BibitemOpen
  \bibfield  {author} {\bibinfo {author} {\bibfnamefont {A.}~\bibnamefont
  {Aghamousa}} \emph {et~al.} (\bibinfo {collaboration} {DESI}),\ }\href@noop
  {} {\  (\bibinfo {year} {2016})},\ \Eprint {http://arxiv.org/abs/1611.00036}
  {arXiv:1611.00036 [astro-ph.IM]} \BibitemShut {NoStop}%
\bibitem [{\citenamefont {Scoccimarro}(2015)}]{Scoccimarro:2015bla}%
  \BibitemOpen
  \bibfield  {author} {\bibinfo {author} {\bibfnamefont {R.}~\bibnamefont
  {Scoccimarro}},\ }\href {\doibase 10.1103/PhysRevD.92.083532} {\bibfield
  {journal} {\bibinfo  {journal} {Phys. Rev. D}\ }\textbf {\bibinfo {volume}
  {92}},\ \bibinfo {pages} {083532} (\bibinfo {year} {2015})},\ \Eprint
  {http://arxiv.org/abs/1506.02729} {arXiv:1506.02729 [astro-ph.CO]}
  \BibitemShut {NoStop}%
\bibitem [{\citenamefont {Philcox}\ \emph {et~al.}(2021)\citenamefont
  {Philcox}, \citenamefont {Hou},\ and\ \citenamefont
  {Slepian}}]{Philcox:2021hbm}%
  \BibitemOpen
  \bibfield  {author} {\bibinfo {author} {\bibfnamefont {O.~H.~E.}\
  \bibnamefont {Philcox}}, \bibinfo {author} {\bibfnamefont {J.}~\bibnamefont
  {Hou}}, \ and\ \bibinfo {author} {\bibfnamefont {Z.}~\bibnamefont
  {Slepian}},\ }\href@noop {} {\  (\bibinfo {year} {2021})},\ \Eprint
  {http://arxiv.org/abs/2108.01670} {arXiv:2108.01670 [astro-ph.CO]}
  \BibitemShut {NoStop}%
\bibitem [{\citenamefont {Schlegel}\ \emph {et~al.}(2019)\citenamefont
  {Schlegel} \emph {et~al.}}]{Schlegel:2019eqc}%
  \BibitemOpen
  \bibfield  {author} {\bibinfo {author} {\bibfnamefont {D.~J.}\ \bibnamefont
  {Schlegel}} \emph {et~al.},\ }\href@noop {} {\  (\bibinfo {year} {2019})},\
  \Eprint {http://arxiv.org/abs/1907.11171} {arXiv:1907.11171 [astro-ph.IM]}
  \BibitemShut {NoStop}%
\bibitem [{\citenamefont {Blas}\ \emph {et~al.}(2011)\citenamefont {Blas},
  \citenamefont {Lesgourgues},\ and\ \citenamefont {Tram}}]{Blas:2011rf}%
  \BibitemOpen
  \bibfield  {author} {\bibinfo {author} {\bibfnamefont {D.}~\bibnamefont
  {Blas}}, \bibinfo {author} {\bibfnamefont {J.}~\bibnamefont {Lesgourgues}}, \
  and\ \bibinfo {author} {\bibfnamefont {T.}~\bibnamefont {Tram}},\ }\href
  {\doibase 10.1088/1475-7516/2011/07/034} {\bibfield  {journal} {\bibinfo
  {journal} {JCAP}\ }\textbf {\bibinfo {volume} {1107}},\ \bibinfo {pages}
  {034} (\bibinfo {year} {2011})},\ \Eprint {http://arxiv.org/abs/1104.2933}
  {arXiv:1104.2933 [astro-ph.CO]} \BibitemShut {NoStop}%
\bibitem [{\citenamefont {Audren}\ \emph {et~al.}(2013)\citenamefont {Audren},
  \citenamefont {Lesgourgues}, \citenamefont {Benabed},\ and\ \citenamefont
  {Prunet}}]{Audren:2012wb}%
  \BibitemOpen
  \bibfield  {author} {\bibinfo {author} {\bibfnamefont {B.}~\bibnamefont
  {Audren}}, \bibinfo {author} {\bibfnamefont {J.}~\bibnamefont {Lesgourgues}},
  \bibinfo {author} {\bibfnamefont {K.}~\bibnamefont {Benabed}}, \ and\
  \bibinfo {author} {\bibfnamefont {S.}~\bibnamefont {Prunet}},\ }\href
  {\doibase 10.1088/1475-7516/2013/02/001} {\bibfield  {journal} {\bibinfo
  {journal} {JCAP}\ }\textbf {\bibinfo {volume} {1302}},\ \bibinfo {pages}
  {001} (\bibinfo {year} {2013})},\ \Eprint {http://arxiv.org/abs/1210.7183}
  {arXiv:1210.7183 [astro-ph.CO]} \BibitemShut {NoStop}%
\bibitem [{\citenamefont {Brinckmann}\ and\ \citenamefont
  {Lesgourgues}(2019)}]{Brinckmann:2018cvx}%
  \BibitemOpen
  \bibfield  {author} {\bibinfo {author} {\bibfnamefont {T.}~\bibnamefont
  {Brinckmann}}\ and\ \bibinfo {author} {\bibfnamefont {J.}~\bibnamefont
  {Lesgourgues}},\ }\href {\doibase 10.1016/j.dark.2018.100260} {\bibfield
  {journal} {\bibinfo  {journal} {Phys. Dark Univ.}\ }\textbf {\bibinfo
  {volume} {24}},\ \bibinfo {pages} {100260} (\bibinfo {year} {2019})},\
  \Eprint {http://arxiv.org/abs/1804.07261} {arXiv:1804.07261 [astro-ph.CO]}
  \BibitemShut {NoStop}%
\bibitem [{\citenamefont {Lewis}(2019)}]{Lewis:2019xzd}%
  \BibitemOpen
  \bibfield  {author} {\bibinfo {author} {\bibfnamefont {A.}~\bibnamefont
  {Lewis}},\ }\href@noop {} {\  (\bibinfo {year} {2019})},\ \Eprint
  {http://arxiv.org/abs/1910.13970} {arXiv:1910.13970 [astro-ph.IM]}
  \BibitemShut {NoStop}%
\bibitem [{\citenamefont {Bernardeau}\ \emph {et~al.}(2002)\citenamefont
  {Bernardeau}, \citenamefont {Colombi}, \citenamefont {Gaztanaga},\ and\
  \citenamefont {Scoccimarro}}]{Bernardeau:2001qr}%
  \BibitemOpen
  \bibfield  {author} {\bibinfo {author} {\bibfnamefont {F.}~\bibnamefont
  {Bernardeau}}, \bibinfo {author} {\bibfnamefont {S.}~\bibnamefont {Colombi}},
  \bibinfo {author} {\bibfnamefont {E.}~\bibnamefont {Gaztanaga}}, \ and\
  \bibinfo {author} {\bibfnamefont {R.}~\bibnamefont {Scoccimarro}},\ }\href
  {\doibase 10.1016/S0370-1573(02)00135-7} {\bibfield  {journal} {\bibinfo
  {journal} {Phys. Rept.}\ }\textbf {\bibinfo {volume} {367}},\ \bibinfo
  {pages} {1} (\bibinfo {year} {2002})},\ \Eprint
  {http://arxiv.org/abs/astro-ph/0112551} {arXiv:astro-ph/0112551 [astro-ph]}
  \BibitemShut {NoStop}%
\bibitem [{\citenamefont {Lazeyras}\ \emph {et~al.}(2016)\citenamefont
  {Lazeyras}, \citenamefont {Wagner}, \citenamefont {Baldauf},\ and\
  \citenamefont {Schmidt}}]{Lazeyras:2015lgp}%
  \BibitemOpen
  \bibfield  {author} {\bibinfo {author} {\bibfnamefont {T.}~\bibnamefont
  {Lazeyras}}, \bibinfo {author} {\bibfnamefont {C.}~\bibnamefont {Wagner}},
  \bibinfo {author} {\bibfnamefont {T.}~\bibnamefont {Baldauf}}, \ and\
  \bibinfo {author} {\bibfnamefont {F.}~\bibnamefont {Schmidt}},\ }\href
  {\doibase 10.1088/1475-7516/2016/02/018} {\bibfield  {journal} {\bibinfo
  {journal} {JCAP}\ }\textbf {\bibinfo {volume} {1602}},\ \bibinfo {pages}
  {018} (\bibinfo {year} {2016})},\ \Eprint {http://arxiv.org/abs/1511.01096}
  {arXiv:1511.01096 [astro-ph.CO]} \BibitemShut {NoStop}%
\end{thebibliography}%

\newpage 

\pagebreak
\widetext
\begin{center}
\textbf{\large Supplemental Material}
\end{center}
\setcounter{equation}{0}
\setcounter{figure}{0}
\setcounter{table}{0}
\setcounter{page}{1}
\makeatletter
\renewcommand{\theequation}{S\arabic{equation}}
\renewcommand{\thefigure}{S\arabic{figure}}
\renewcommand{\bibnumfmt}[1]{[S#1]}
\renewcommand{\citenumfont}[1]{S#1}

\section{Details of the theory model}

Following the notation of \cite{Chudaykin:2020aoj}, the $Z_2$ kernel necessary to derive $P_{12}(\mathbf{k})$ in Eq.~\eqref{eq:P12_expression} is 
\begin{equation}
\begin{split}
        Z_2(\k_1,\k_2)  &=\frac{b_2}{2}+b_{\mathcal{G}_2}\bigg(\frac{(\k_1\cdot \k_2)^2}{k_1^2k_2^2}-1\bigg)
+b_1 F_2(\k_1,\k_2)+f\mu^2 G_2(\k_1,\k_2) \\
&\;\;\;\;+\frac{f\mu k}{2}\bigg(\frac{\mu_1}{k_1}(b_1+f\mu_2^2)+
\frac{\mu_2}{k_2}(b_1+f\mu_1^2)
\bigg)\,\,,
\end{split}
\end{equation}
where $F_2$ and $G_2$ are the SPT kernels \cite{Bernardeau:2001qr} 
\begin{equation}
\begin{split}
F_2(\k_1,\k_2) &= \frac{5}{7} + \frac{2}{7}\frac{(\k_1\cdot \k_2)^2}{k_1^2k_2^2} + \frac{1}{2}\frac{\k_1\cdot \k_2}{k_1k_2}\bigg(\frac{k_1}{k_2} + \frac{k_2}{k_1}\bigg)\,\,, \\
G_2(\k_1,\k_2) &= \frac{3}{7} + \frac{4}{7}\frac{(\k_1\cdot \k_2)^2}{k_1^2k_2^2} + \frac{1}{2}\frac{\k_1\cdot \k_2}{k_1k_2}\bigg(\frac{k_1}{k_2} + \frac{k_2}{k_1}\bigg) \,\,,
\end{split}
\end{equation}
and $\mu_i = \hat{\mathbf{k}}_i\cdot\hat{\mathbf{z}}$. Our non-Gaussian model depends on the quadratic bias  
$b_2$ and the tidal bias $b_{\mathcal{G}_2}$, which also enter the tree-level 
bispectrum model. 
For dark matter halos,
these biases have the 
following approximate 
dependence on the 
linear bias $b_1$:
\be 
\label{eq:coev}
b_{\mathcal{G}_2}=-\frac{2}{7}(b_1-1)\,\,,\quad b_2 = 
0.412 - 2.143 b_1 + 0.929 b_1^2 + 0.008 b_1^2 - \frac{8}{21}(b_1 - 1)\,\,.
\ee 
The relationship 
$b_{\mathcal{G}_2}(b_1)$
follows from 
the Lagrangian local 
in matter density model~\cite{Desjacques:2016bnm}, while the function $b_2(b_1)$
is fit from the halo
simulation data~\cite{Lazeyras:2015lgp}.
These relationships, strictly 
speaking, do not 
hold for simulated 
galaxies, but the 
deviations are not significant 
given the precision
of the current data~\cite{Eggemeier:2021cam,Ivanov:2021kcd}. We use the relations~\eqref{eq:coev}
in our aggressive analysis, evaluating 
them for the best-fit values 
of $b_1$ taken from the 
baseline results. 

\section{Effects of PNG on galaxy power spectra and bispectra}

\begin{figure}[ht!]
    \centering
    \includegraphics[width=0.49\textwidth]{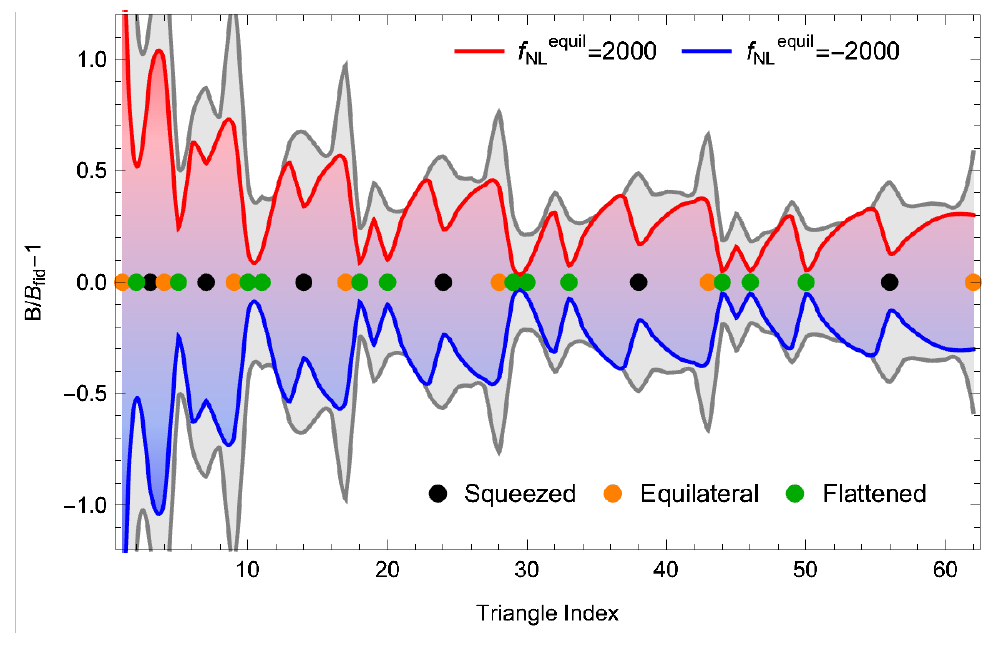}
     \includegraphics[width=0.49\textwidth]{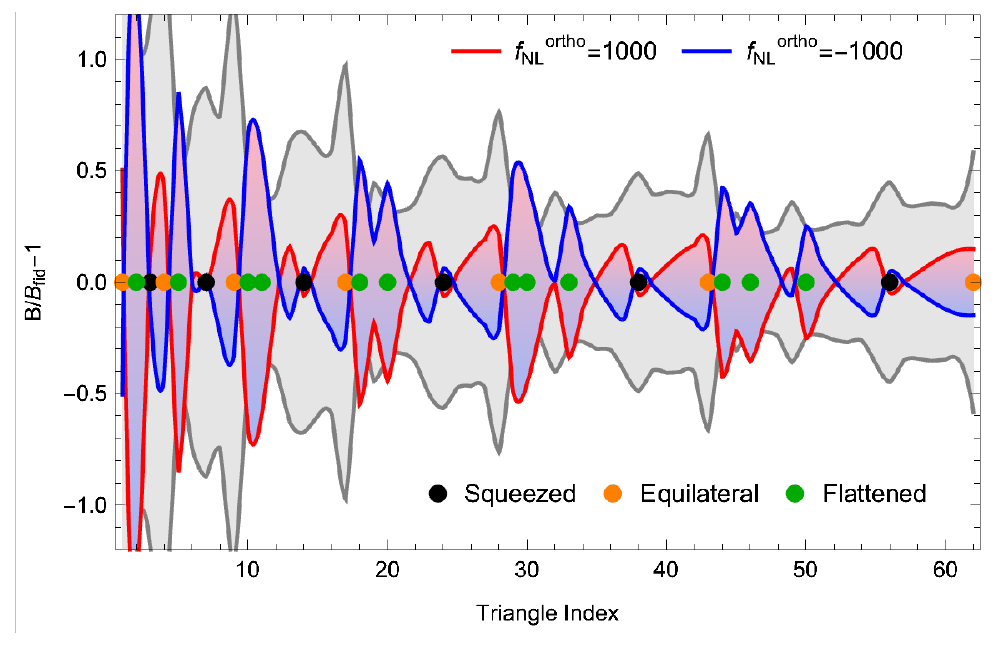}
      \includegraphics[width=0.49\textwidth]{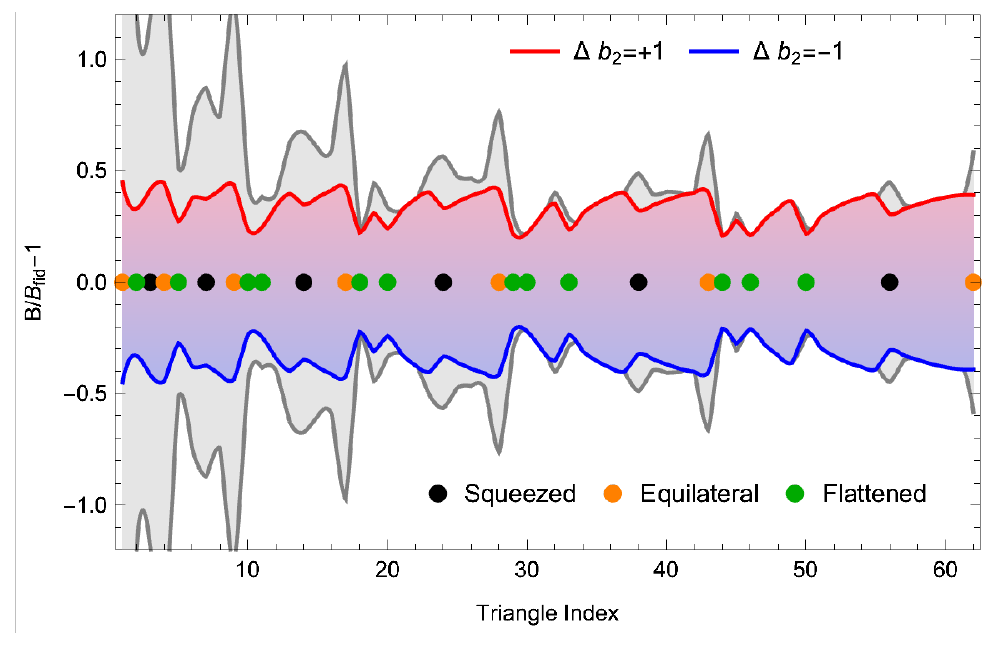}
       \includegraphics[width=0.49\textwidth]{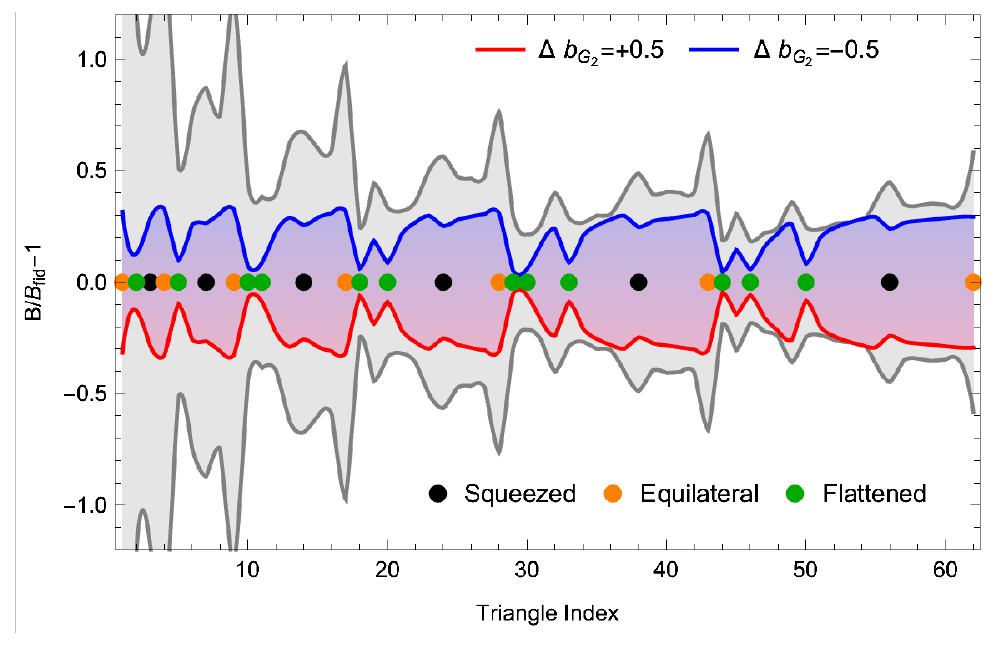}
    \caption{Residual variations of the galaxy bispectrum monopole
    w.r.t. variations of $\fnl^{\rm equil}$ (upper left panel), $\fnl^{\rm ortho}$ (upper right panel), quadratic bias $b_2$ (lower left panel), and tidal bias
    $b_{\mathcal{G}_2}$ (lower right panel). Black, orange and green
    dots denote the squeezed ($k_3=0.015~\hMpc$, $k_2=k_1$), equilateral ($k_1=k_2=k_3$), and flattened ($k_2=k_3$, $2k_2=k_1+0.005~\hMpc$)
    triangle configurations, respectively. 
    These variations 
    have distinctive shape dependence 
    and therefore can be constrained 
    by the data.}
    \label{fig:residual}
\end{figure}

In this Section we
present plots that give some 
intuition on the source of our $f_{\rm NL}$ constraints. 
In Fig.~\ref{fig:residual}
we illustrate the effect of variations
of the parameters $\fnl^{\rm equil},\fnl^{\rm ortho},b_2,b_{\mathcal{G}_2}$
on the galaxy bispectrum monopole $B$. Parameters are varied within the following 
intervals:
\be 
\label{eq:intervals}
\fnl^{\rm equil}\in [-2000,2000]\,\,,\quad 
\fnl^{\rm ortho}\in [-1000,1000]\,\,,\quad 
b_2\in [b_2^{\rm fid}-1,b_2^{\rm fid}+1]\,\,,\quad
b_{\mathcal{G}_2}\in [b_{\mathcal{G}_2}^{\rm fid}-0.5,b_{\mathcal{G}_2}^{\rm fid}+0.5]\,\,,
\ee 
where $b_2^{\rm fid},b^{\rm fid}_{\mathcal{G}_2}$
are fiducial parameters equal to the best-fit values.
We present the residuals $B/B_{\rm fid.}-1$ as a function 
of the triangle index  
of our bispectrum data vector, where $B_{\rm fid.}$
is the best-fit bispectrum model
for the NGC z3 ($z=0.61$) data
chunk, obtained in the analysis 
with $\fnl^{\rm equil}=0,\fnl^{\rm ortho}=0$.
Each triangle is labelled by 
its bin center wavenumbers ($k_1,k_2,k_3$), whose magnitudes are constrained to be $k\in [0.015,0.08)~\hMpc$.
Indices corresponding 
to squeezed, equilateral,
and flattened configurations
are denoted by black, orange, and green
dots in the $x$ axis, and a shaded gray contour on the background shows the error bars 
of the NGC z3 data. We chose 
the extremal values of the 
varied 
parameters 
in Eq.~\eqref{eq:intervals}
such that the
resulting residual curve 
has 
a $1\sigma$ deviation at least 
for several data points. This suggests 
that 
the cumulative $\chi^2$ should be significant, 
and the corresponding 
variations should be constrained
by the data.

The first important observation
is that the parameters of interest
have very different shape dependence.
$\fnl^{\rm equil}$ mostly enhances 
equilateral triangles ($k_1=k_2=k_3$); the effect on the 
squeezed triangles ($k_3=0.015\,\hMpc$, $k_1=k_2$, $k_1>k_3$) is somewhat weaker,
whilst the flattened triangles
($k_3=k_2$, $2k_2= k_1+0.005\,\hMpc$)
are affected very little. 
In contrast, $\fnl^{\rm ortho}$
peaks
sharply at the flattened and equilateral triangles. Note that these effects 
are of the opposite sign, which
is a known signature 
of the orthogonal template. 
Remarkably, the effect 
on the squeezed triangles 
is vanishingly small. 

Now let us discuss the quadratic 
bias parameters, which represent
a late-time non-Gaussian contribution. The effect of $b_2$ is almost uniform
for all triangles. 
In contrast, $b_{\mathcal{G}_2}$
noticeably amplifies the squeezed
and equilateral triangles,
while the flattened 
triangles are almost unaffected. 

All in all, we see that the 
data should distinguish 
between the shapes of interest. 
The qualitative picture of that is as follows. 
$\fnl^{\rm ortho}$ is the only shape
that sharply peaks at the flattened triangles.
In addition, it produces 
specific ``opposite'' peaks 
in the equilateral triangles. 
Clearly, this pattern cannot be 
reproduced by any other parameter 
and hence we do not expect 
$\fnl^{\rm ortho}$ to be degenerate 
with them. 
The signature of $b_2$ is also quite unique,
as it is the only shape that enhances
all triangles in a uniform way.
Hence, it should be robustly constrained.
The only noticeable 
degeneracy we expect is between 
$\fnl^{\rm equil}$ and $b_{\mathcal{G}_2}$,
because both corresponding shapes
mostly enhance squeezed and equilateral 
configurations, while leaving 
the flattened triangles almost intact. 
The degeneracy between these gets broken 
by the relative height of the squeezed 
and equilateral triangles:
$b_{\mathcal{G}_2}$ enhances both these
triangles uniformly, 
while $\fnl^{\rm equil}$
generates some contrast by amplifying 
more 
the equilateral configurations. 

\begin{figure}[ht!]
    \centering
    \includegraphics[width=0.49\textwidth]{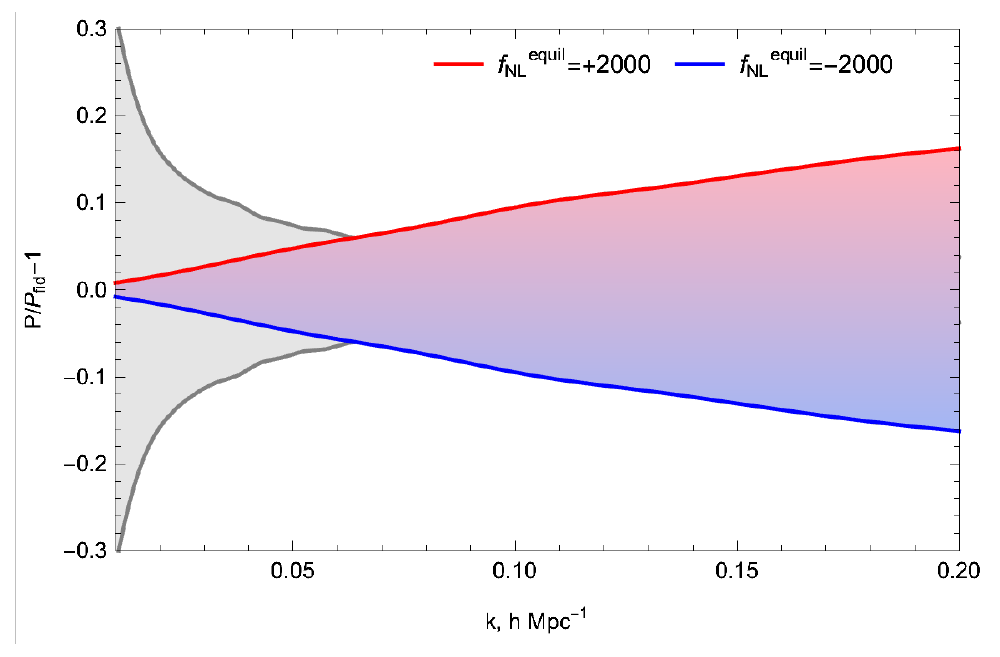}
     \includegraphics[width=0.49\textwidth]{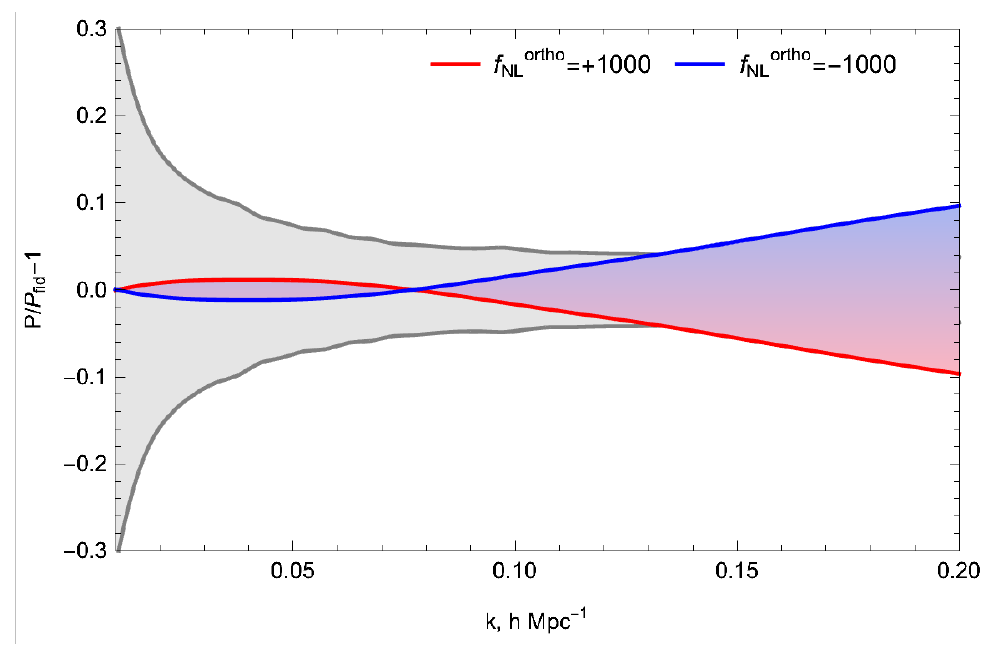}
      \includegraphics[width=0.49\textwidth]{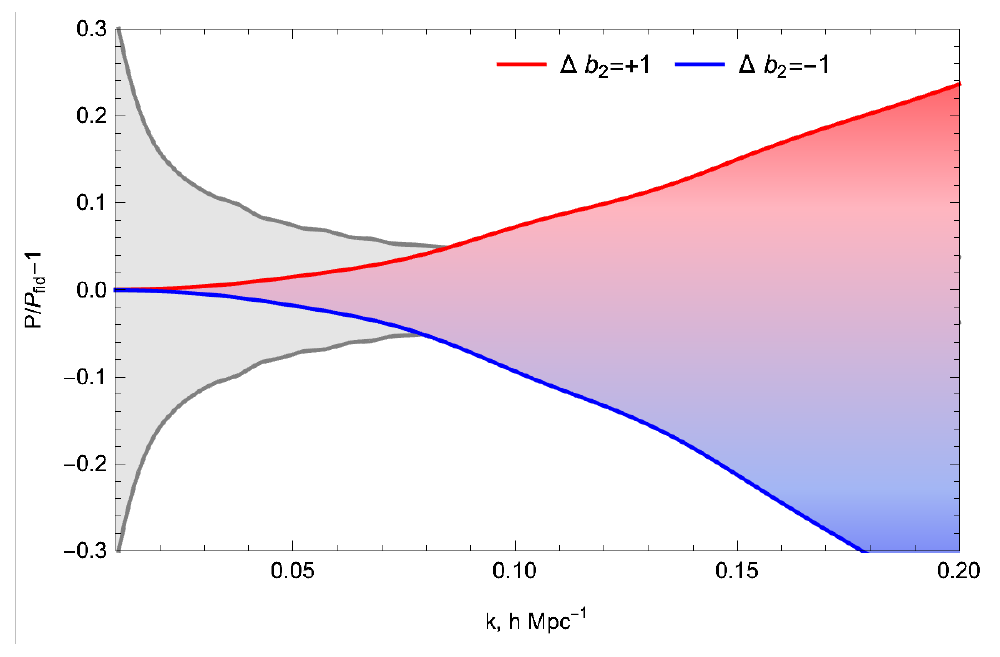}
       \includegraphics[width=0.49\textwidth]{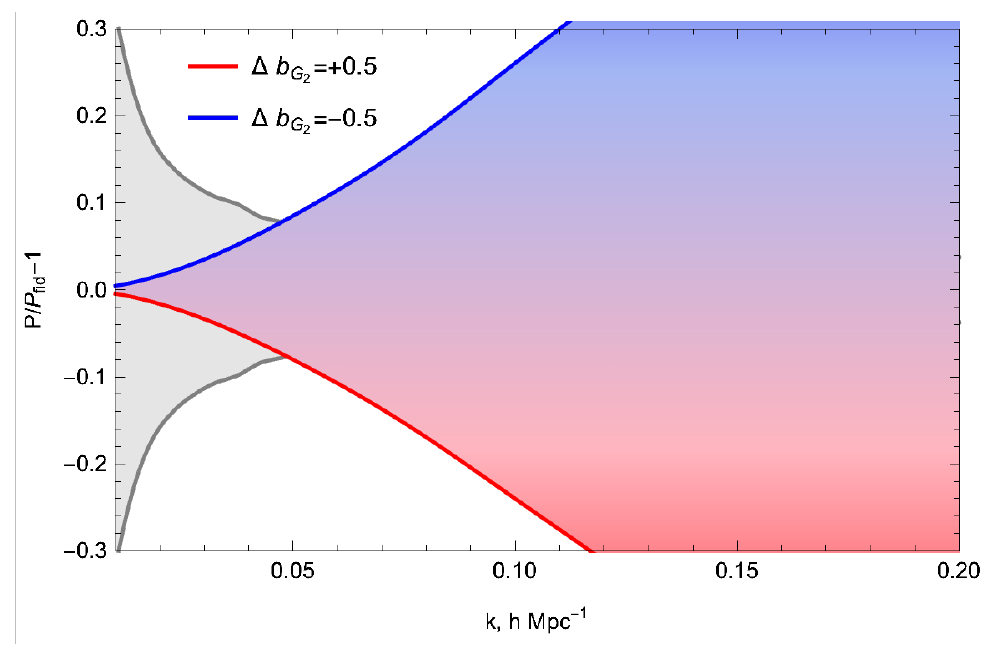}
    \caption{Residual variations of the galaxy power spectrum monopole $P_0$
    w.r.t. variations of $\fnl^{\rm equil}$ (upper left panel), $\fnl^{\rm ortho}$ (upper right panel), quadratic bias $b_2$ (lower left panel), and tidal bias
    $b_{\mathcal{G}_2}$ (lower right panel).}
    \label{fig:resP}
\end{figure}

In Fig.~\ref{fig:resP} we repeat the same
exercise for the galaxy power
spectrum monopole $P$
for the wavenumbers 
relevant for our analysis, $k\in [0.01,0.2)\,\hMpc$. 
We focus on the $P_{12}$-type matter loop 
corrections only, because the effects 
of the linear PNG bias clearly cannot
be constrained without a prior on $b_\zeta$.
Looking at Fig.~\ref{fig:resP}, we see that, although power spectrum modulations
produced by 
$\fnl^{\rm equil}\sim 1000$,
$\fnl^{\rm ortho}\sim 1000$
are nominally larger than 
the error bars, the signal
produced by these shapes is a very 
smooth function of scale. 
Thus, the PNG terms are quite degenerate
with the Gaussian non-linear corrections, explaining why we cannot get 
strong constraints 
on NLPNG from the power
spectra alone. For very large $\fnl^{\rm equil},\fnl^{\rm ortho}$, however, one can break the
degeneracies, i.e.\ using the non-monotonic behaviour 
of the $\fnl^{\rm ortho}$ term 
on large scales $\lesssim 0.1\,\hMpc$.
This explains why in the mock simulation 
data analysis we were able to break degeneracies
for very large $\fnl^{\rm ortho}\sim 1000$.
However, the total volume 
of this simulation
was $40$ times larger than the actual BOSS
survey volume. This demonstrates 
that the galaxy 
power spectrum alone does not
have enough power 
to provide any interesting 
constraints on PNG, though it is still
instrumental in breaking parameter
degeneracies
relevant for the bispectrum,
e.g. the 
notorious degeneracy 
between $b_1$
and $\sigma_8$.

\section{Full triangle plots and constraint tables}

\begin{figure}
    \centering
    \includegraphics[width=\textwidth]{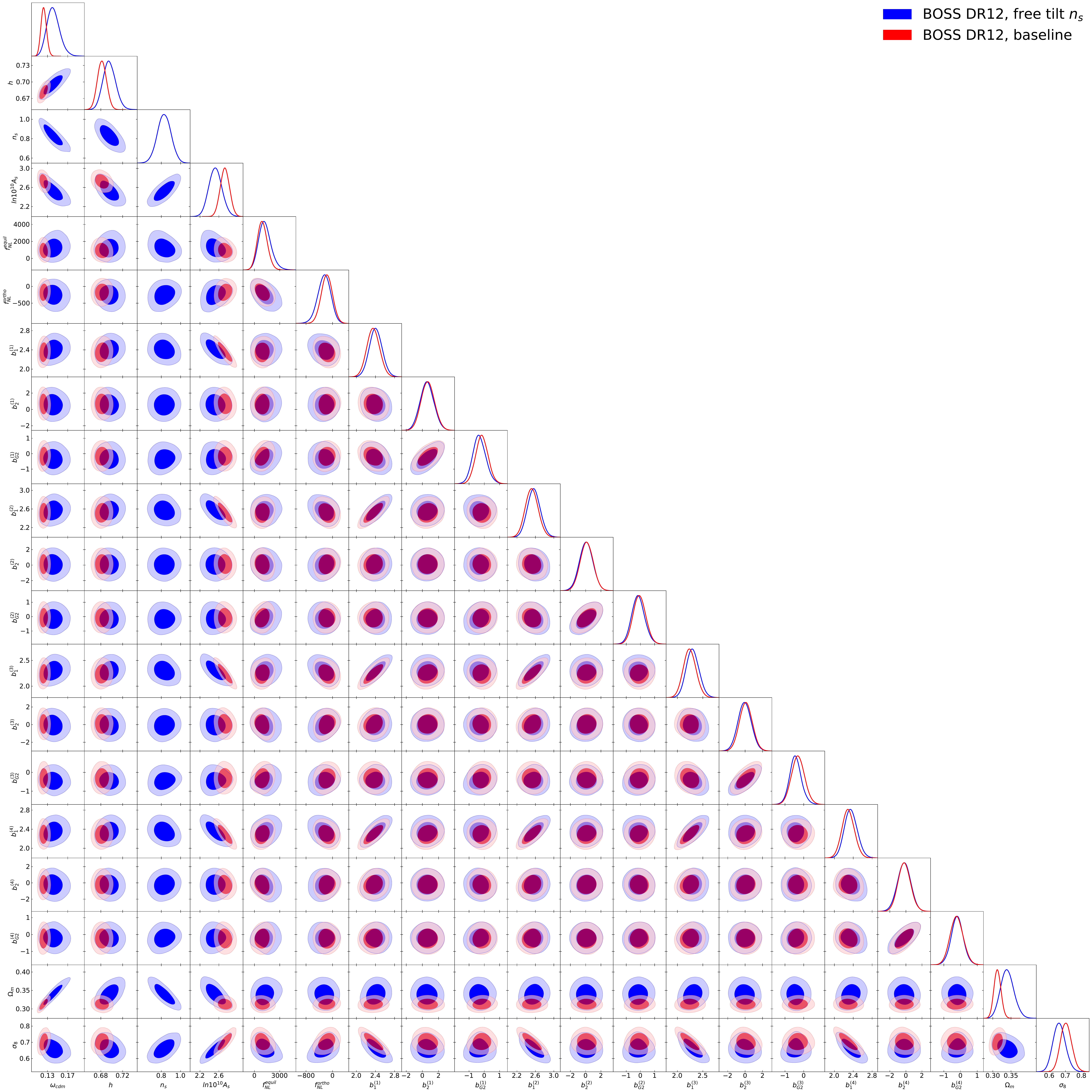}
    \caption{Full posterior plot of the cosmological and nuisance parameter posteriors measured from the BOSS DR12 data (with a BBN prior on the $\omega_b$), with 
    the 
    free power spectrum tilt $n_s$ (blue contours) and with $n_s$ fixed to the \textit{Planck} prior (red contours, baseline choice). The corresponding parameter constraints are given in Tab.~\ref{tab: all-constraints}.}
    \label{fig: all-corner}
\end{figure}

In this Section we present a full triangle plot (Fig.~\ref{fig: all-corner}) 
and parameter constraint table (Tab.~\ref{tab: all-constraints})  from two PNG analyses
of BOSS DR12: (a) with free power
spectrum tilt $n_s$ and (b) with 
a \textit{Planck} prior on $n_s$.
The latter is our baseline choice. 
We present results for the joint analysis of the full power spectrum, BAO, and bispectrum datasets.

\begin{table}[ht!]
    \centering
   \rowcolors{2}{white}{vlightgray}
  \begin{tabular}{|c|cccc|cccc|}
 \hline
 & \multicolumn{4}{c|}{BOSS DR12 + free $n_s$} & \multicolumn{4}{c|}{BOSS DR12, baseline}\\\hline
\quad \textbf{Parameter}\quad\quad & best-fit & mean$\,\pm\,\sigma$ & \quad 95\% lower \quad & \quad 95\% upper \quad & best-fit & mean$\,\pm\,\sigma$ & \quad 95\% lower \quad & \quad 95\% upper \quad \\ \hline
$\omega_{cdm }$ &$0.1262$ & $0.1413_{-0.015}^{+0.011}$ & $0.1159$ & $0.1682$ &$0.1197$ & $0.1225_{-0.0059}^{+0.0053}$ & $0.1115$ & $0.1339$ \\
$h$ &$0.6848$ & $0.6954_{-0.013}^{+0.011}$ & $0.6719$ & $0.7195$ &$0.6827$ & $0.682_{-0.0088}^{+0.0085}$ & $0.665$ & $0.6992$ \\
$\mathrm{ln}\left(10^{10}A_{s}\right)$ & $2.739$ & $2.533_{-0.15}^{+0.14}$ & $2.249$ & $2.826$ &$2.716$ & $2.728_{-0.1}^{+0.1}$ & $2.528$ & $2.929$ \\
$n_{s }$ &$0.9324$ & $0.8323_{-0.074}^{+0.075}$ & $0.6858$ & $0.9801$& - & - & - & - \\
$10^{-2}f^{\rm equil}_{{\rm NL} }$ &$1.82$ & $12.39_{-8.6}^{+6.7}$ & $-3.179$ & $28.49$ & $2.278$ & $9.421_{-6.5}^{+5.7}$ & $-2.779$ & $21.91$  \\
$10^{-2}f^{\rm ortho}_{{\rm NL} }$ &$-0.0628$ & $-2.484_{-1.9}^{+2.1}$ & $-6.557$ & $1.484$ & $-0.7874$ & $-1.732_{-1.7}^{+1.8}$ & $-5.265$ & $1.756$ \\
$b^{(1)}_{1 }$ & $2.323$ & $2.404_{-0.14}^{+0.14}$ & $2.127$ & $2.687$ &$2.408$ & $2.353_{-0.14}^{+0.13}$ & $2.078$ & $2.63$ \\
$b^{(1)}_{2 }$ &$0.5561$ & $0.5731_{-0.9}^{+0.84}$ & $-1.159$ & $2.295$ & $0.9517$ & $0.675_{-0.88}^{+0.83}$ & $-1.012$ & $2.396$ \\
$b^{(1)}_{{\mathcal{G}_2} }$ &$-0.3561$ & $-0.324_{-0.46}^{+0.41}$ & $-1.197$ & $0.57$ &$-0.1242$ & $-0.1568_{-0.42}^{+0.41}$ & $-0.9911$ & $0.6793$ \\
$b^{(2)}_{1 }$ &$2.502$ & $2.566_{-0.15}^{+0.14}$ & $2.282$ & $2.858$ &$2.563$ & $2.521_{-0.15}^{+0.14}$ & $2.239$ & $2.809$  \\
$b^{(2)}_{2 }$ &$0.08728$ & $0.07405_{-0.91}^{+0.88}$ & $-1.685$ & $1.883$ &$0.4217$ & $0.1337_{-0.88}^{+0.87}$ & $-1.617$ & $1.86$ \\
$b^{(2)}_{{\mathcal{G}_2} }$ &$-0.2795$ & $-0.1499_{-0.49}^{+0.43}$ & $-1.079$ & $0.8198$ &$-0.1956$ & \,\,$-0.06784_{-0.47}^{+0.45}$ & $-0.9817$ & $0.8616$ \\
$b^{(3)}_{1 }$ &$2.198$ & $2.299_{-0.14}^{+0.12}$ & $2.041$ & $2.561$ &$2.26$ & $2.242_{-0.13}^{+0.12}$ & $1.99$ & $2.499$ \\
$b^{(3)}_{2 }$ &$-0.03962$ & \,\,$-0.05736_{-0.78}^{+0.77}$ & $-1.631$ & $1.503$ &$0.2868$ & $0.08216_{-0.77}^{+0.75}$ & $-1.428$ & $1.617$ \\
$b^{(3)}_{{\mathcal{G}_2} }$ &$-0.4604$ & $-0.4348_{-0.33}^{+0.3}$ & $-1.087$ & $0.2549$ &$-0.3913$ & $-0.3014_{-0.37}^{+0.35}$ & $-1.022$ & $0.4259$ \\
$b^{(4)}_{1 }$ &$2.223$ & $2.348_{-0.15}^{+0.13}$ & $2.073$ & $2.636$ &$2.317$ & $2.293_{-0.14}^{+0.13}$ & $2.025$ & $2.564$ \\
$b^{(4)}_{2 }$ &$-0.04403$ & $-0.2223_{-0.84}^{+0.83}$ & $-1.913$ & $1.461$ &$0.01938$ & $-0.1926_{-0.83}^{+0.77}$ & $-1.79$ & $1.433$\\
$b^{(4)}_{{\mathcal{G}_2} }$ &$-0.1569$ & $-0.1981_{-0.37}^{+0.37}$ & $-0.9765$ & $0.5782$ &$-0.3865$ & $-0.2292_{-0.41}^{+0.39}$ & $-1.028$ & $0.5808$  \\\hline
$\Omega_{m }$ &$0.3183$ & $0.3394_{-0.02}^{+0.018}$ & $0.3022$ & $0.3778$ &$0.3062$ & $0.3129_{-0.01}^{+0.0096}$ & $0.2932$ & $0.333$ \\
$H_0$ &$68.48$ & $69.54_{-1.3}^{+1.1}$ & $67.19$ & $71.95$ &$69.44$ & $69.57_{-1.3}^{+1.1}$ & $67.19$ & $72.03$ \\
$\sigma_8$ &$0.7124$ & $0.663_{-0.042}^{+0.038}$ & $0.5841$ & $0.7452$ &$0.689$ & $0.7039_{-0.038}^{+0.034}$ & $0.6335$ & $0.7753$ \\
\hline
 \end{tabular}
 \caption{Full parameter constraints from the analysis of full BOSS DR12 data including the bispectrum, combined with a BBN prior on the $\omega_b$ (both columns), and an additional Planck prior on $n_s$ (right columns). We present the best-fit values, the mean, 68\%, and 95\% confidence level results in each case, and show the derived parameters in the bottom rows. The superscripts on bias parameters indicate the sample, in the order NGC z3, SGC z3, NGC z1, SGC z1. The associated two-dimensional posteriors are shown in Fig.~\ref{fig: all-corner}.}\label{tab: all-constraints}
\end{table}
 
\end{document}